\documentclass[aps,pra,reprint,superscriptaddress,showpacs]{revtex4-1}
\usepackage{graphicx}
\usepackage{amssymb}
\usepackage{amsmath}
\usepackage{bm}
\usepackage{verbatim}
\usepackage[colorlinks=true,urlcolor=blue,citecolor=blue,linkcolor=blue]{hyperref}

\usepackage{enumitem} 

\newcommand{\e}{\epsilon}

\newcommand{\mk}[1]{\hat\sigma^{#1}}
\newcommand{\mn}[1]{\hat\tau^{#1}}

\newcommand{\bea}{\begin{eqnarray}}
\newcommand{\eea}{\end{eqnarray}}
\newcommand{\mathsym}[1]{{}}
\newcommand{\unicode}[1]{{}}
\newcommand{\be}{\begin{equation}}
\newcommand{\ee}{\end{equation}}
\newcommand{\bemm}{\begin{multline}}
\newcommand{\enmm}{\end{multline}}
\newcommand{\ben}{\begin{equation*}}
\newcommand{\een}{\end{equation*}}
\newcommand{\bmmn}{\begin{multline*}}
\newcommand{\emmn}{\end{multline*}}

\newcommand{\ov}[1]{\overline{#1}}
\newcommand{\ovm}[1]{\hat{\overline{#1}}}

\newcommand{\bra}{\left\langle}
\newcommand{\ket}{\right\rangle}
\newcounter{ourcount}
\usepackage{epigraph}

\begin{document}

\title{Quasiparticle relaxation in superconducting nanostructures }
\author{Yahor Savich}
\affiliation{School of Physics and Astronomy, University of Minnesota, Minneapolis, MN 55455, USA}
\author{Leonid Glazman}
\affiliation{Department of Physics, Yale University, New Heaven, CT 06520, USA}
\author{Alex Kamenev}
\affiliation{School of Physics and Astronomy, University of Minnesota, Minneapolis, MN 55455, USA}
\affiliation{William I. Fine Theoretical Physics Institute, University of Minnesota, Minneapolis, MN 55455, USA}

\date{\today}
\vspace{0.1cm}

\begin{abstract}
We examine energy relaxation of non-equilibrium quasiparticles in ``dirty'' superconductors with the electron mean free path much shorter than the superconducting coherence length. Relaxation of low-energy non-equilibrium quasiparticles is dominated by phonon emission. We derive the corresponding collision integral and find the quasiparticle relaxation rate. The latter is sensitive to the breaking of time reversal symmetry (TRS) by a magnetic field (or magnetic impurities). 
As a concrete application of the developed theory, we address quasiparticle trapping by a vortex and a current-biased constriction. We show that trapping of hot quasiparticles may predominantly occur at distances from the vortex core, or the constriction, significantly exceeding the superconducting coherence length.
\end{abstract}

\maketitle


\section{Introduction}

Current interest to the dynamics of Bogoliubov quasiparticles in superconductors is motivated in no small part by the efforts aimed at building a quantum computer. The actively explored ``cavity QED'' architecture relies on quantum coherence of qubits built of conventional superconductors~\cite{MHDRJS2013}. Realization of the topological quantum computing requires coherence of devices made of proximitized semiconductor quantum wires brought into the p-wave superconducting state by applied magnetic field~\cite{Karzig2017}. In any of the concepts, the presence of quasiparticles is detrimental to the coherence. The Q-factors of the parts comprising the cavity-QED qubit are reduced by quasiparticles. They also are able to ``poison'' the Majorana states, which are central for the topological quantum computing.

The development of the qubit technology has advanced also the ability to monitor the quasiparticles population and dynamics. Time-resolved measurements performed with the transmon~\cite{Chen2014} and fluxonium quibits~\cite{Pop2014,Vool2014} allowed the experimentalists to measure the rates of quasiparticle trapping by a single vortex in a superconducting strip,  to identify minute dissipative currents of quasiparticles across a Josephson junction (thus resolving a longstanding ``$\cos\varphi$-problem''~\cite{Barone}), and to monitor the spontaneous temporal variations of the quasiparticle density.

Measurements~\cite{Pop2014,Vool2014} did confirm that at low temperatures (less than $\sim 0.1T_c$ of Al) the quasiparticle density, albeit low, far exceeds the equilibrium value. Furthermore, statistics of temporal variations of the density substantially differs from thermal noise. Sources of excess quasiparticles remain unknown, and planting quasiparticle traps~\cite{Rajauria2009,Rajauria2012,Riwar2017} remains a viable way of improving the device performance. Trap is a spatial region with a suppressed value of superconducting gap. Suppression may be achieved, e.g.,  through the proximity effect, or through local violation  of time-reversal invariance (as it naturally happens in and around the core of a vortex). Energy loss in the trap (mostly due to phonon emission) prevents a quasiparticle from exiting into the region with the nominal gap value.

The importance of the quasiparticle energy relaxation in device applications, and the newly acquired ability of precise measurements~\cite{Chen2014,Pop2014,Vool2014} of the quasiparticles dynamics prompts us to revisit the kinetic theory of quasiparticles interacting with phonons in a disordered superconductor. We derive the corresponding collision integrals and relaxation rates which then may be used in sophisticated phenomenological models of quasiparticle diffusion and trapping~\cite{Riwar2017,Ullom2012}.

In considering the electron-phonon interaction in disordered metals, we follow the seminal works of Tsuneto~\cite{Tsuneto} and Schmid~\cite{Schmid},  who established the correct form of the electron-phonon interaction in the limit of short electron mean free path, $ql\ll 1$ (here $\bf q$ is the phonon wave vector). We incorporate the electron-phonon interaction in the general framework of Keldysh non-linear sigma-model. It allows us to consider on equal footing normal metals and superconductors, and it becomes especially convenient for describing the effect of breaking the time-reversal symmetry (TRS). 

The kernel of the collision integral for electrons in normal metal which we find in the unified technique, agrees with the earlier results~\cite{Reizer,Yudson} obtained diagrammatically; this kernel depends only on the energy transferred in the collision to a phonon. Considering the Bogoliubov quasiparticles, we are able to cast the result for the collision integral in the conventional terms of the quasiparticle energy distribution functions. In the presence of TRS, the corresponding kernel factorizes on two terms: the normal-state kernel and a combination of the Bogoliubov transformation parameters.  Factorization takes place also if TRS is broken; in that case, the second factor is determined by the proper solution of the Usadel equation. In either of the two cases, the second factor depends separately on the initial and final energy of a quasiparticle.

The additional (compared to the normal state) energy dependence of the kernel affects the dependence of the quasiparticle relaxation rate on its energy. These rates, in turn, determine the effectiveness of trapping. As an example of application of the developed theory, we consider trapping of a quasiparticle by an isolated vortex and a current-biased constriction. In both cases there is a pattern of super-currents, slowly decaying as a function of distance, $\sim r^{-1}$, from the vortex core, or $\sim r^{1-d}$ from a constriction with $d$-dimensional superconducting leads. These super-currents lead to a weak breaking of TRS and thus suppression of the energy gap and modification of the energy dependence of the density of states (DOS). Such a suppression allows quasiparticles to be trapped already very far from the vortex core or the constriction.  For low enough phonon temperature and relatively ``hot''  quasiparticles this peripheral shallow trapping proves to be more efficient than the deep trapping by the core of the vortex, or the constriction. We discuss possible relation of the theory to experiments \cite{Chen2014,Siddiqi2014}.    

The paper is organized as follows:   in Section \ref{sec:el-ph} we review the theory of electron-phonon interactions in disordered normal metals. We derive the corresponding Keldysh non-linear sigma-model and use it to obtain the 
electron-phonon collision integral in the dirty limit. In Section \ref{sec:disordered-superconductors}  we generalize the sigma-model on superconductors, including those with broken TRS, and in Section \ref{sec:kinetics} derive kinetic equation for the quasiparticle distribution.
Section \ref{sec:trapping} is devoted to applications of the theory to  trapping by vortex and current-biased constriction as well as discussion of the existing experiments. We summarize with a brief discussion in Section \ref{sec:discussion}.   
Two Appendices present an alternative derivation of the sigma-model and summarize results for ultrasound attenuation.

\section{Electron-phonon interactions in disordered normal metals} 
\label{sec:el-ph}

\subsection{Interaction vertex}
\label{sec:interaction-vertex}

Theory of electron-phonon interactions in normal disordered metals had a long and, at times, controversial history. Early considerations were based on the Fr\"ohlich Hamiltonian \cite{Frohlich}, which  assumes screened Coulomb interactions between electron density and induced lattice charge, $e\rho_0\,\mathrm{div}\, {\mathbf{u}}$, created by phonon displacement $\mathbf{u}(\mathbf{r},t)$. Here $e\rho_0$ is the uniform lattice charge density. Due to global neutrality it is exactly equal to the electron density:      
\begin{equation}
							\label{eq:bare density}
\rho_0=\int\limits^{p_F} \frac{d^d \mathbf{p}}{(2\pi)^d} = \int\limits_0^{\e_F} d\e \,\nu(\e) = \frac{v_F p_F \nu}{d},
\end{equation} 
where normal metal DOS is $\nu(\e)=(\e/\e_F)^{d/2-1}\nu$ and $\nu=\nu(\e_F)$. While perfectly legitimate in the clean case, the Fr\"ohlich Hamiltonian misses an important piece of the physics in the ``dirty'' limit $ql\ll 1$, where $q$ is phonon wavenumber and $l$ is electron elastic mean free path. 

As was first realized by Pippard \cite{Pippard}, phonons not only deform the lattice, but also displace impurities, transforming formerly static impurity potential $U_\mathrm{imp}(\mathbf{r})$ into the dynamic one, $U_\mathrm{imp}(\mathbf{r})\to U_\mathrm{imp}(\mathbf{r}+\mathbf{u}(\mathbf{r},t))$. Colloquially,  this leads to the electron density being dragged along with the lattice displacement and providing a perfect compensation for the induced lattice charge  $e\rho_0\,\mathrm{div}\, {\mathbf{u}}$. 
In other words,  the displaced impurity potential provides fast elastic relaxation of the electron distribution around the Fermi surface locally deformed by phonons. These ideas were put on a quantitative basis by Tsuneto \cite{Tsuneto} and Schmid \cite{Schmid}, who showed that  in the limit $ql\ll 1$  the Fr\"ohlich Hamiltonian should be substituted by another 
effective electron-phonon interaction vertex: 
\begin{equation}
 							\label{eq:electron-phonon} 
iS_{e:ph}\!=\!\!\int\!\! dt \sum\limits_\mathbf{p,q} \bar\psi\left(\mathbf{p}+\frac{\mathbf{q}}{2},t\right)\! \Gamma_{\mu\nu}(\mathbf{p}) i\mathbf{q}^\mu \mathbf{u}^\nu_{\mathbf{q},t} \psi\left(\mathbf{p}-\frac{\mathbf{q}}{2},t\right)\!, 
\end{equation}
where $\bar\psi$ and $\psi$ are electrons creation and annihilation operators  and $\Gamma_{\mu\nu}(\mathbf{p}) $ is the traceless tensor 
\begin{equation}
					\label{eq:Gamma}
\Gamma_{\mu\nu}(\mathbf{p}) = \mathbf{p}_\mu  \mathbf{v}_\nu - \frac{p_Fv_F}{d}\, \delta_{\mu\nu}.
\end{equation}
Notice that, in view of Eq.~(\ref{eq:bare density}), the last term here represents the Fr\"ohlich coupling $-\nu^{-1} (\rho_0\mathrm{div}\, {\mathbf{u}})(\bar\psi\psi)$. Upon averaging over the Fermi surface it is exactly compensated by the first term in Eq.~(\ref{eq:Gamma}). The remaining coupling is of a quadrupole nature, as seen from Eqs.~(\ref{eq:electron-phonon}),   (\ref{eq:Gamma}). This leads to a significantly weaker electron-phonon coupling, than the one inferred from the Fr\"ohlich term \cite{Altshuler1978}. A number of subsequent studies \cite{Reizer,Yudson,Shtyk} reaffirmed  validity of the Schmid coupling  (\ref{eq:electron-phonon}), (\ref{eq:Gamma}) from various perspectives. 

The most straightforward way to derive Eqs.~(\ref{eq:electron-phonon}), (\ref{eq:Gamma}) \cite{Tsuneto,Shtyk-thesis} is by performing a unitary transformation, which yields a Hamiltonian in the co-moving reference frame, where the impurity potential is static. We shall nor repeat this derivation here. Instead, we accept  Eqs.~(\ref{eq:electron-phonon}), (\ref{eq:Gamma}) as a starting point and derive an effective non-linear sigma model which incorporates electron-phonon interaction in the Schmid form. In Appendix \ref{sec:alternative}  we provide an alternative derivation of the sigma-model, which proceeds in the laboratory reference frame and deals with a dynamic random potential $U_\mathrm{imp}(\mathbf{r}+\mathbf{u}(\mathbf{r},t))$. We show that it brings the same effective sigma model, justifying the use of the effective electron-phonon vertex in the Schmid form~(\ref{eq:electron-phonon}), (\ref{eq:Gamma}).

\subsection{Non-linear sigma model}
\label{sec:sigma-model}

We now perform the standard \cite{Andreev1999,Kamenev2011} averaging over the  static disorder and introduce the non-local field $Q_{t,t'}(\mathbf{r})$ to split emerging four-fermion term.   
The resulting action, including electron-phonon coupling, Eqs.~(\ref{eq:electron-phonon}), (\ref{eq:Gamma}), is now quadratic in the fermionic fields which may be integrated out in the usual way, leading to 
\begin{equation}
				\label{eq:tr-log-1}
iS=  -\frac{\pi\nu}{4\tau}\,\mathrm{Tr}\{ Q^2\} + \mathrm{Tr}\log\left\{ G_0^{-1} +\frac{i}{2\tau}Q + \Gamma_{\mu\nu} \partial^\mu \mathbf{u}^\nu  \right\}, 
\end{equation}
where the inverse {\em bare} electron Green function is given by $G_0^{-1}=i\partial_t +\nabla^2/2m+\mu\approx i\partial_t + i\mathbf{v}_\mu\partial^\mu$.    

From this point on, one proceeds along the standard root of deriving Keldysh non-linear sigma-model \cite{Andreev1999, Kamenev2011}. To this end one passes to the Keldysh $2\times 2$ structure, by splitting the contour on forward and backward branches and performing 
Keldysh rotation. Upon this procedure the fields acquire the matrix structure, e.g.  $\mathbf{u}\to \hat{\mathbf{u}}=\mathbf{u}^\alpha\hat\gamma^\alpha$, where $\alpha=cl,q$ denotes classical and quantum Keldysh components and $\hat\gamma^{cl}=\hat\sigma^0, \hat\gamma^{q}=\hat\sigma^1$ are the two vertex matrices in the Keldysh space. 

One then realizes that the soft diffusive modes of the action are described by the manifold $\hat Q^2 =1$ and therefore one can write $\hat Q= \hat{\cal R}^{-1} \hat\Lambda\hat{\cal R}$, where $\hat\Lambda$ is the Green function in coinciding spatial points,
\begin{equation}
					\label{eq:Lambda}
\hat\Lambda=\frac{i}{\pi\nu}\sum_{\mathbf{p}}\hat G_0(\mathbf{p},\epsilon)=
\left(\begin{array}{cc} 1& 2F_\epsilon\\ 0&-1\end{array} \right), 
\end{equation}
and $ F_\epsilon$ is a distribution function. Rotation matrices, $\hat{\cal R}^{-1}$, belong to an appropriate symmetry group.   
One then introduces {\em dressed} Green function $\hat G =\left(\hat G_0^{-1} +\frac{i}{2\tau}\hat \Lambda\right)^{-1}$ and rewrites the action (\ref{eq:tr-log-1}) as 
\begin{equation}
				\label{eq:tr-log-2}
iS= \mathrm{Tr}\log\!\left\{\! 1\!+\! \hat G\hat{\cal R}[\hat G_0^{-1}\!, \hat{\cal R}^{-1}] +\hat G\,\hat{\cal R}\,  \Gamma_{\mu\nu}\, \partial^\mu \hat{\mathbf{u}}^\nu \,  \hat{\cal R}^{-1}\!  \right\}\!. 
\end{equation}
Finally, one expands the logarithm here to the lowest  non-vanishing orders. This way one obtains the standard non-linear sigma-model action (first neglecting electron-phonon $\Gamma$-term):
\begin{equation}  
				\label{eq:S0}                             
iS_{\hat Q}=  -
\frac{\pi\nu}{4}\,
\mathrm{Tr}\big\{ D\,({{\partial}}_{\mathbf{r}}
{\hat Q})^{2}-4   \partial_{t}{\hat Q}   \big\} , 
\end{equation}
where $D=v_F^2\tau/d$ is the diffusion constant and $d$ is the dimensionality of electron system.  We focus now on the  phonon-induced  term.  It is easy to see that the first order in  $ \Gamma$ term vanishes due to the fact that the integral over the Fermi surface $\int d\Omega_\mathbf{p}  \Gamma_{\mu\nu}(\mathbf{p}) =0$. It is this point, where the Schmid coupling, Eqs.~(\ref{eq:electron-phonon}), (\ref{eq:Gamma}), is qualitatively different from the Fr\"olich one (the latter would bring the first order $\mathrm{Tr}\big\{(\rho_0\mathrm{div}\, \hat{\mathbf{u}})\hat Q\big\}$ term). Going to the second order in $\Gamma$, one finds:
\begin{equation}
iS_{\hat Q,\mathbf{u}}= -\frac 1 2\, \mathrm{Tr}\big\{ \hat G\,\hat{\cal R}\,  \Gamma_{\mu\nu}\, \partial^\mu \hat{\mathbf{u}}^\nu \,  \hat{\cal R}^{-1}  
\hat G\,\hat{\cal R}\,  \Gamma_{\eta\lambda}\, \partial^\eta \hat{\mathbf{u}}^\lambda \,  \hat{\cal R}^{-1} \big\}. 
\end{equation}
We use now 
\begin{equation}
\hat G_\mathbf{p}=\frac 1 2 \,G^R_\mathbf{p}\,(1+\hat \Lambda) + \frac 1 2\, G^A_\mathbf{p}\,(1-\hat \Lambda), 
\end{equation}
along with
\begin{eqnarray}
&&\sum\limits_\mathbf{p} \!\mathbf{p}_{\mu} \mathbf{v}_\nu   G^R_\mathbf{p}\, \mathbf{p}_\eta \mathbf{v}_\lambda  G^A_\mathbf{p}\\
&&=\frac{2\pi\nu\tau p_F^2v_F^2}{d(d+2)}\left( \delta_{\mu\nu}\delta_{\eta\lambda} + \delta_{\mu\eta}\delta_{\nu\lambda} + \delta_{\mu\lambda}\delta_{\nu\eta}\right) \nonumber
\end{eqnarray}
to find
\begin{equation}
				\label{eq:S-u}
iS_{\hat Q,\mathbf{u}}= \frac{\pi\nu D\, p_F^2}{4}\,\,  \mathrm{Tr}\big\{ [\hat Q\,,\partial^\mu \hat{\mathbf{u}}^\nu]  [\hat Q\,,\partial^\eta \hat{\mathbf{u}}^\lambda] \big\}\, \Upsilon_{\mu\nu,\eta\lambda}, 
\end{equation}
where
\begin{equation}
				\label{eq:upsilon}
\Upsilon_{\mu\nu,\eta\lambda} = \frac{1}{d+2}\left[\delta_{\mu\eta}\delta_{\nu\lambda}  + \delta_{\mu\lambda}\delta_{\nu\eta} -\frac{2}{d}\, \delta_{\mu\nu}\delta_{\eta\lambda}\right].
\end{equation}
The local vertex (\ref{eq:S-u}) is the leading term describing interaction of phonons with the electronic degrees of freedom in disordered metals, in  $ql\ll 1$ limit. The naive deformation potential term $S\propto \rho_0 \mathrm{Tr}\{\hat Q \,\mathrm{div}\, \hat{\mathbf{u}} \}$ is absent due to the perfect screening manifested in the traceless form of the electron-phonon vertex (\ref{eq:Gamma}). See also Appendix \ref{sec:alternative} for more discussion of this issue. The second order cross-term between the two terms in the logarithm in Eq.~(\ref{eq:tr-log-2}) leads to $ S\propto \rho_0 \tau D \mathrm{Tr}\{\nabla^2 \hat Q\, \mathrm{div}\, \hat{\mathbf{u}} \}$. It is of the order $(ql)\ll 1$ of the leading term  (\ref{eq:S-u}) and thus should not be kept within the accuracy of the adopted approximations. 

The effective electron-phonon sigma model, Eqs.~(\ref{eq:S0}) and (\ref{eq:S-u}), should be supplemented with the standard phonon action. In the Keldysh technique it is given by 
\begin{equation}
				\label{eq:phonon-action}
iS_\mathbf{u} =i\frac{\rho_m}{2}\! \sum\limits_{\mathbf{q},\omega,j}\!  
\bar{\mathbf{u}}^{\mu,\alpha}_{\mathbf{q},\omega} 
\!\left [\omega^2 -\left( \omega_\mathbf{q}^{(j)}\right)^2\right] \hat\sigma^1_{\alpha\beta} \eta_{\mu\nu}^{(j)}(\mathbf{q}) 
\, \mathbf{u}^{\nu,\beta}_{\mathbf{q},\omega},  
\end{equation}
where $\rho_m$ is the material mass density, $j=l,t$ labels longitudinal and transversal polarizations encoded by the projectors
\begin{equation}
			\label{eq:polarizations}
\eta_{\mu\nu}^{(l)}(\mathbf{q})= \frac{q_\mu q_\nu}{q^2};\qquad \eta_{\mu\nu}^{(t)}(\mathbf{q})= \delta_{\mu\nu} -\frac{q_\mu q_\nu}{q^2}, 
\end{equation}
and $\omega_\mathbf{q}^{(j)} =v_j q$ is the acoustic phonons dispersion with the speed of sound $v_{l,t}$. Indexes 
$\alpha,\beta=cl,q$ and Pauli matrix $\hat\sigma^1$ act in the $2\times 2$ Keldysh space. We will need an imaginary part of the corresponding retarded propagator: 
\begin{eqnarray}
			\label{eq:ph-propagator}
&&\mathrm{Im}\, U_{\nu\mu}^{R}(\mathbf{q},\omega) = \mathrm{Re}\, \langle \mathbf{u}^{\nu,cl}_{\mathbf{q},\omega} 
\bar{\mathbf{u}}^{\mu,q}_{\mathbf{q},\omega}\rangle 	\\
&&= \sum\limits_j
\frac{ \eta_{\nu\mu}^{(j)}(\mathbf{q})}{\rho_m}\, \frac{\pi }{2 \omega_\mathbf{q}^{(j)}} \left[\delta\left(\omega-\omega_\mathbf{q}^{(j)}\right) - \delta\left(\omega+\omega_\mathbf{q}^{(j)}\right)\right]. \nonumber
\end{eqnarray}
The corresponding Keldysh component is given by the fluctuation-dissipation relation: $U_{\nu\mu}^{K}={\cal B}_\omega(U_{\nu\mu}^{R}-U_{\nu\mu}^{A})$, where ${\cal B}_\omega=\coth (\omega/2T)$ is the bosonic distribution function.  

The effective action, Eqs. (\ref{eq:S0}), (\ref{eq:S-u}), and (\ref{eq:phonon-action}) with the vertices defined in Eqs.~(\ref{eq:upsilon}) and (\ref{eq:polarizations}) serves as the starting point for investigating the kinetics of electrons and phonons. We relegate the evaluation of the ultrasonic attenuation to Appendix \ref{app:attenuation}, where we reaffirm the known results~\cite{Schmid,Abrikosov,Shtyk,Shtyk-thesis} obtained by different techniques, and proceed to study the electron kinetics.

\subsection{Electron-phonon collision integral}

To derive collision integral for electron-phonon interactions one  first integrates over the phonon displacements $\mathbf{u}(\mathbf{r},t)$ with the help of Eq.~(\ref{eq:ph-propagator}) 
to obtain the collision action from Eq.~(\ref{eq:S-u}):
\begin{equation}
				\label{eq:S-collisions-normal}
S_{\mathrm{coll}}\!=\! \frac{\pi\nu D p_F^2}{4}\,  \mathrm{Tr}\big\{ \hat Q_{\e-\omega,\e'-\omega} \hat\gamma^{\alpha}\hat Q_{\e',\e} \hat\gamma^{\beta}  \big\}\,U^{\alpha\beta}_{\nu\lambda}\mathbf{q}^\mu \mathbf{q}^\eta \Upsilon_{\mu\nu,\eta\lambda}, 
\end{equation}
where  $U^{\alpha\beta}_{\nu\lambda}=U^{\alpha\beta}_{\nu\lambda}(\mathbf{q},\omega)$ and summation over $\e,\e',\omega, \mathbf{q}$ are understood.  One now looks for the stationary point equation for the action $S_Q+S_{\mathrm{coll}}$.  It's Keldysh (1,2) component constitutes the kinetic equation \cite{Andreev1999,Kamenev2011} for  the distribution function $F_\e$ in Eq.~(\ref{eq:Lambda}), 
\begin{eqnarray} 
				\label{eq:kinetic-normal}
 \partial_t F_{\e} - \nabla_\mathbf{r}\left[D\nabla_\mathbf{r} F_\e \right] = -2 I_\mathrm{coll}[ F_\epsilon(\mathbf{r},t)], 
 \end{eqnarray}
where the collision integral is given by  
\be
                  \label{coll_def}
I_\mathrm{coll}[ F_\epsilon(\mathbf{r},t)]  =-\frac{1}{2\pi\nu}\, \bra\left(\frac{\delta iS_\mathrm{coll}}{\delta \hat Q_{\e\e}(\mathbf{r})}\right)^{(1,2)}\ket_{\hat Q}. 
\ee
The variational derivative here ought to be restricted to the 
sigma-model target space, $\hat Q^2=1$. A way to insure this is to use parameterization  
$\hat Q \to e^{-\hat W/2}\hat Q e^{\hat W/2}\approx\hat Q+\frac 12 [\hat Q,\hat W]$ and expand the action  (\ref{eq:S-collisions-normal}) to the linear order in  $\hat W_{\e\e}$.  Here $\hat W$'s are infinitesimal  generators of the symmetry transformations. Because of the local nature of the vertex in Eq.~(\ref{eq:S-collisions-normal}), the $\hat Q$ integration may be substituted by the stationary point: $\hat Q\to \hat \Lambda$. This way one obtains:
\begin{eqnarray}
                                                 \label{eq:collisions-normal}
&& I_\mathrm{coll}[ F_\epsilon] = i\frac{ D p_F^2}{8} \sum\limits_{\mathbf{q},\e' }  \left[\hat\gamma^{\beta}  \hat\Lambda_{\epsilon'} \hat\gamma^{\alpha}  \hat\Lambda_{\epsilon} - \hat\Lambda_{\epsilon} \hat\gamma^{\beta}  \hat\Lambda_{\epsilon'}\hat\gamma^{\alpha} 
\right]^{(1,2)}\nonumber \\
&&\times U^{\alpha\beta}_{\nu\lambda}(\mathbf{q},\e-\e') \mathbf{q}^\mu \mathbf{q}^\eta \Upsilon_{\mu\nu,\eta\lambda} = 
\frac14\sum\limits_{\e' }  M_{\e,\e'} \,\mathcal{I}[F]; \\
 &&\mathcal{I}[F]=-1+F_\e F_{\e'}+{\cal B}_{\e-\e'}[F_\e-F_{\e'}] \nonumber \,.
\end{eqnarray}
The phonon matrix element in the collision integral (\ref{eq:collisions-normal}) in normal metal is found to depend only on the energy difference, $M_{\e,\e'}=M^N_{\e-\e'}$, where
\begin{equation}
							\label{eq:matrix-element-normal}
M_\omega^N\! =\! 2Dp_F^2\!  \sum\limits_{\mathbf{q}}  \mathrm{Im}\!\left[U^{R}_{\nu\lambda}(\mathbf{q},\omega)\right]  \mathbf{q}^\mu \mathbf{q}^\eta \Upsilon_{\mu\nu,\eta\lambda} \!=\! M^{(l)}_\omega\! + \! M^{(t)}_\omega\!,
\end{equation}
and the superscripts $j=l,t$ stand for longitudinal and transverse modes, respectively. Expressing the fermion and boson distributions in terms of the respective occupation numbers, $F_\epsilon=1-2f_\epsilon$ and ${\cal B}_{\omega}=1+ 2N_\omega$, we may bring $I_\mathrm{coll}[F_\epsilon]$ to the standard ``in minus out'' form,
\begin{equation}
I_\mathrm{coll}[ f_\epsilon]=\!\int\!\frac{d\e'}{2\pi}\, M_\omega^N
[N_{\omega} f_{\e'}(1-f_{\e}) - (1+N_{\omega}) f_\e(1-f_{\e'})]\,;
								\label{null}
\end{equation}
where $\omega=\e-\e'$ and $I_\mathrm{coll}[ f_\epsilon]=0$ in equilibrium.

Employing Eqs.~(\ref{eq:upsilon}), (\ref{eq:ph-propagator}) and (\ref{eq:matrix-element-normal}) one finds:
\begin{equation}
				\label{eq:matrix-element-normal-1}
M^{(j)}_\omega \!= \! \frac{b_j\pi Dp_F^2}{\rho_m} \sum\limits_{\mathbf{q}} \frac{q^2}{\omega_\mathbf{q}^{(j)}} \,\delta\!\left(\omega-\omega_\mathbf{q}^{(j)}\!\right)\!=\! \frac{b_j\pi\Omega_{d}}{(2\pi)^d}\,  \frac{Dp_F^2 \omega^d}{\rho_m v_j^{d+2}},
\end{equation}
where $\Omega_d=2\pi^{d/2}/\Gamma(d/2)$ is the area of  $S_{d-1}$ unit sphere and $d$ is the effective phonon dimensionality.  The coefficients are given $b_t= (d-1)/(d+2)$; $b_l=2b_t/d$ (in these instances $d$ is the dimensionality of the electron system).   

These results for the normal metal were derived in Ref.~\cite{Schmid,Reizer,Yudson} using diagrammatic techniques. Here we reproduced them through the sigma-model technique, which is much more suitable for treating the superconducting case, considered below.  We notice that $M_\omega^{(l)}$ of Eq.~(\ref{eq:matrix-element-normal-1}) is factor $(v_l/v_F)^2\ll 1$ smaller than that of Ref.~\cite{Altshuler1978} for $\omega<v_l^2/D$. The latter was obtained  with the Fr\"ohlich coupling (i.e. disregarding impurities shifting with the lattice deformations).  The two approaches give comparable results for the longitudinal phonons at $\omega\approx v_l/l$ where they both match with the clean limit expectation $M^{(l)}_\omega \propto v_F p_F^2\omega^{d-1}/(\rho_m v_l^{d+1})$. The transversal phonons give the dominant contribution to the collision integral in the disordered limit $\omega < v_t/l$, since typically $v_t < v_l$. However, in the opposite -- clean limit, the transversal matrix element is  $M^{(t)}_\omega \propto Dp_F^2\omega/(\rho_m v_t^3 l^2)$, \cite{Yudson} and is less important than  the longitudinal one.

For comparison, the electron-electron collision integral may be written in the form of Eq.~(\ref{null}), with the bosonic occupation number ${N}_\omega=\omega^{-1}\int d\e'' f_{\e''}(1-f_{\e''-\omega})$ and a different matrix element given by:
$M^N_\omega \to M^{e:e}_\omega\propto \omega^{d/2-1}D^{-d/2}/\nu$, \cite{Altshuler-Aronov}. As a result the ratio of electron-electron and phonon matrix elements in normal metals is
\begin{equation}
\frac{M^{e:e}_\omega}{M^N_\omega}\propto \frac{M}{m} \left(\frac{v_j^2}{ D\omega}\right)^{d/2+1}\propto 
\left(\frac{m}{ M}\right)^{d/2}\left(\frac{1}{ \omega\tau}\right)^{d/2+1},
\end{equation}
where $M$ is the ion mass and we used that  $\rho_m \propto M\nu p_F^2/m$ and $v_j^2\propto v_F^2m/M$. 
Therefore electron-electron relaxation in normal metals dominates for the energy transfer $\omega < \tau^{-1}(m/M)^{d/(d+2)}$.

\section{Disordered superconductors} 
\label{sec:disordered-superconductors} 

\subsection{Sigma-model}

The non-linear sigma-model is readily extended to  disordered superconductors  \cite{Skvortsov,Kamenev2011}. It is written in terms of the local pair correlation function $\check Q_{t,t'}(\mathbf{r})\propto \langle \Psi(\mathbf{r},t)\Psi^\dagger(\mathbf{r},t')\rangle$, where $\Psi(\mathbf{r},t)$ is the four component spinor in the Nambu and Keldysh subspaces. As a result $\check Q_{t,t'}(\mathbf{r})$ is  a $4\times 4$ matrix, as well as the matrix in the time, $t,t'$, space.   It satisfies the non-linear condition $\check Q^2=1$. Its dynamics is governed by the  action:
\begin{equation}  
				\label{eq:S0-supercond}                             
iS_{\check Q:\check \Delta}=  -
\frac{\pi\nu}{8}\,
\mathrm{Tr}\big\{ D\,({{\partial}}_{\mathbf{r}}
{\check Q})^{2}-4  \check{\cal{T}}_3  \partial_{t}{\check Q}   + 4i
\check{\Delta}  {\check Q}  \big\}, 
\end{equation}
where $\check\Delta(\mathbf{r},t)=\Delta(\mathbf{r},t)\gamma^{cl}\otimes\hat\tau^{+} - \overline\Delta(\mathbf{r},t)\gamma^{cl}\otimes\hat\tau^{-}$ is the order parameter matrix.  To discuss broken TRS later on, we have also included a vector potential through the long derivative: 
\begin{equation}
\label{eq:derivative}
{{\partial}}_{\mathbf{r}}{\check Q}={{\nabla}}_{\mathbf{r}}{\check Q} + i[\mathbf{A} \check {\cal{T}}_3,\check Q]. 
\end{equation}
Hereafter  $\hat\tau^{0,1,2,3}$ are Pauli matrices in the Nambu space and $\check{\cal{T}}_3 = \hat\gamma^{cl}\otimes\hat\tau^3$. Here the operation $\mathrm{Tr}$ involves trace in $4\times 4$ Nambu-Keldysh space, as well as trace in time (or equivalently energy) space and the spatial integration. 

The electron-phonon interactions are  given by Eq.~(\ref{eq:S-u})  (with factor $1/2$ to compensate for the Nambu doubling of the degrees of freedom),  where the displacement field $\check{\mathbf{u}}$ is proportional to $\hat\tau^0$ matrix in the Nambu space.  The corresponding collision action, obtained by integrating out the phonon degrees of freedom, is given by Eq.~(\ref{eq:S-collisions-normal}) (again with factor $1/2$).  Its variation over $\check Q$ leads to the collision integral in the form of Eq.~(\ref{eq:collisions-normal}). The major difference of the superconducting case is that  the $\check \Lambda_\e$ matrices in Eq.~(\ref{eq:collisions-normal}) are rotated in Nambu space, as explained below.

Taking variation of the effective action (\ref{eq:S0-supercond}), (\ref{eq:S-u}) with respect to the $\check Q$ as explained after Eq.~(\ref{coll_def}), one obtains the saddle point Usadel equation~\cite{Usadel,Kamenev2011}
\begin{equation}
							\label{SuperCond-Usadel}
\big\{\check{\cal{T}}_3\partial_{t},{\check{Q}}\big\}_+ - \hat{{\partial}}_{\mathbf{r}}\big(D {\check{Q}}\,
\hat{{\partial}}_{\mathbf{r}} {\check{Q}}\big)- 
i\big[ \check{\Delta}, {\check{Q}}\big]=\frac{1}{\pi\nu} \frac{\delta S_\mathrm{coll}}{ \delta \check Q},
\end{equation}
We look for a solution of this equation $\check Q=\check \Lambda$ in the standard form respecting causality:  
\begin{equation}
                                               \label{SuperCond-classical-ansatz}
{\check \Lambda}=\left(\begin{array}{cc}\hat{\Lambda}^{R}
& \hat{\Lambda}^{K}
\\ 0 & \hat{\Lambda}^{A} \end{array}\right)_{\!\!K}\,,
\end{equation}
with retarded, advanced and Keldysh components being matrices in
the Nambu subspace.   The non-linear constraint $ \check \Lambda^2= 1$ is resolved as 
\begin{equation}\label{SuperCond-Normalizations}
\hat{\Lambda}^{R}\hat{\Lambda}^{R}=\hat{\Lambda}^{A}\hat{\Lambda}^{A}=\hat{1}\,,\quad\quad
\hat{\Lambda}^{K}=\hat{\Lambda}^{R} \hat{F}-\hat{F}\hat{\Lambda}^{A}\,, 
\end{equation} 
where $\hat{F}$ is a distribution matrix in the Nambu space, which may be written as \cite{LarkinOvchinnikov}
$\hat{F}=F^L_{\e\e'}(\mathbf{r})\hat\tau^0 + F^T_{\e\e'}(\mathbf{r})\hat\tau^3$. Here $F^{L,T}$ are longitudinal (odd with respect to energy permutation) and transverse (even in energy permutation) components of the quasiparticle distribution function. These two  are responsible for the transport of energy and charge correspondingly.  (The conventional distribution functions $F^{L,T}_\epsilon(\mathbf{r},t))$ are obtained by Wigner transformation with $(\e+\e')/2\to \e$ and $\e-\e'\to t$.) Since the transversal component usually decays fast to zero, we shall  primarily focus only on the long-lived longitudinal component of the non-equilibrium quasiparticle distribution and often  omit the superscript for brevity  $F^L_\e(\mathbf{r},t)=F_\epsilon(\mathbf{r},t)$.   In thermal equilibrium  
$F^L_\e =\tanh \epsilon/2T$, while $F^T_\e =0$. 

The non--linear constraints $(\hat{\Lambda}^{R(A)})^{2}=\hat{1}$, Eq.~(\ref{SuperCond-Normalizations}),  may be explicitly resolved in the Nambu space by the angular
parametrization~\cite{BelzigWilhelm,Kamenev2011}:
\begin{eqnarray}
						\label{eq:angular}
&& \hat{\Lambda}^{R}(\mathbf{r},\epsilon)\!=\! \left(\!\!\begin{array}{cc}
\cosh\vartheta & \sinh\vartheta\,\,  e^{i\chi}\\
-\sinh\vartheta\,\, e^{-i\chi} &     -\cosh\vartheta
\end{array}\!\!\right)_{\!\!N}\!=\! \hat V^{-1} \hat\tau^3 \hat V ; \\
                                                    \label{eq:angular-A}
&& \hat{\Lambda}^{A}(\mathbf{r},\epsilon)\!=\!\left(\!\!\begin{array}{cc}
-\cosh\overline\vartheta & -\sinh\overline\vartheta\,\,  e^{i\overline\chi}\\
\sinh\overline\vartheta\,\, e^{-i\overline\chi} &     \cosh\overline\vartheta
\end{array}\!\right)_{\!\!N}\!=\! -\ovm {V}^{-1}\mn3 \ovm V , \nonumber
\end{eqnarray}
where $\vartheta(\mathbf{r},\epsilon)$ and $\chi(\mathbf{r},\epsilon)$
are {\em complex}, coordinate-- and energy--dependent angles. Here 
\begin{equation}
                                   \label{eq:V}
  \hat V_\epsilon(\mathbf{r}) = e^{\frac{\vartheta}{2}\hat\tau^1}e^{-i\frac{\chi}{2}\hat\tau^3}\,;
  \quad\quad 
  \ovm{V_\e}(\mathbf{r}) = e^{\frac{\ov\vartheta}{2}\mn1}e^{-i\frac{\ov\chi}{2}\mn3},
\end{equation}   
notice that in presence of the phase $\chi$ the matrix $\ovm{V_\e}$ is not a  complex conjugate of $  \hat V_\epsilon$. 
The full saddle point $\check \Lambda$-matrix (\ref{SuperCond-classical-ansatz}), (\ref{SuperCond-Normalizations}) then acquires the form 
\be
                   \label{eq:Utau3U}
\check \Lambda(\mathbf{r},\epsilon) =
\check U^{-1}_\epsilon(\mathbf{r})\, \mn 3 \otimes \mk 3\,  \check U_\epsilon(\mathbf{r}),
\ee 
where $\mk 3$ is the Keldysh space matrix and  
\be
                       \label{eq:U}
\check U=
\begin{pmatrix}
 \hat V &&  \hat{V} \hat F\\
 0 && -{\ovm V}\\
\end{pmatrix}_{\!\!K}; \qquad
\check U^{-1}=
\begin{pmatrix}
 \hat V^{-1} &&\hat F \ovm{V}^{-1}\\
 0 && -{\ovm V}^{-1}\\
 \end{pmatrix}_{\!\!K}.
\ee

The expectation value of the order parameter  satisfy the self-consistency equation, obtained by variation of $S_{\check Q,\check\Delta}-\frac{i\nu}{2\lambda} \mathrm{Tr}\{\check \Delta \hat\sigma^1\otimes\hat\tau^0\check\Delta\} $ over the quantum component $\Delta^q$. This leads to  (we assume $F^T=0$): 
\begin{equation}
                                                                                     \label{eq:self-consistency} 
\Delta =\frac{\lambda}{4}\int\limits_{-\omega_D}^{\omega_D} d\epsilon \, F_\epsilon^L\, 
 \left[ \sinh\vartheta  +\sinh\overline\vartheta \right],   
 \end{equation}
where $\lambda$ is the BCS interaction constant and $\omega_D$ is the Debye frequency cutoff. 

In the absence of the vector potential, i.e. with unbroken TRS, substituting Eqs.~(\ref{eq:angular}) into the retarded and advanced components of the Usadel equation (\ref{SuperCond-Usadel}) one finds for the Nambu angle: 
\be
\label{eq:Usadel-angular}
\epsilon  =  \Delta \coth{\vartheta}  = \Delta \coth{\bar\vartheta}.
\ee   
For $|\epsilon|>\Delta$ one thus finds that $\vartheta(\e)$ is real and 
\be
\label{eq:xi}
\cosh \vartheta =\frac{\epsilon}{\xi_\e};\quad \sinh \vartheta =\frac{\Delta}{\xi_\e};\quad 
\xi_\e\equiv \mathrm{sign}(\epsilon)\sqrt{\epsilon^2-\Delta^2}.
\ee   
Within  the energy gap, $|\epsilon| <\Delta$,  the angle is  $\vartheta = -i\pi/2+\theta$, with real $\theta$.  For all energies the  following symmetry relation holds $\vartheta(-\epsilon) = -\bar\vartheta(\epsilon)$. The local DOS is expressed through the Nambu angle as  
\begin{equation}
\label{eq:DOS}
\nu(\epsilon)= {\nu\over 2}\, \mathrm{Re}\,\mathrm{tr}\big\{\hat \tau^3 \hat \Lambda^R \big\} =  \nu   \mathrm{Re} \cosh\vartheta(\epsilon)= \nu\,\frac{\e}{\xi_\e}\,\Theta(|\e|-\Delta),
\end{equation}
where $\Theta$ is the step function.

\subsection{Superconductors with broken TRS}
\label{sec:broken-TRS}

In many cases of the practical interest the vector potential (and hence the phase $\chi$) changes slowly on the scale of the superconducting coherence length. In these cases one may disregard the gradient terms in the action (\ref{eq:S0-supercond}) and write it as:  
\begin{equation}  
				\label{eq:S0-no-gradients}                             
iS^{(0)}_{\check Q,\check\Delta}=  -
\frac{\pi\nu}{8}\,
\mathrm{Tr}\Big\{ 
-\frac{\gamma}{2}[\check{\cal{T}}_3,{\check Q}]^{2} + 4i \epsilon  \check{\cal{T}}_3  {\check Q}   + 4i
\check{\Delta} {\check Q}  \Big\}, 
\end{equation}
where $\gamma= 2D\mathbf{A}^2$ is the energy scale associated with the local breaking of TRS. 
For a vortex $A=1/(2r)$, where $r$ is distance from the core, and thus $\gamma =\frac 1 2 \Delta(\xi/r)^2$, where $\xi=\sqrt{D/\Delta}$ is the coherence length. For a thin  film of width $d<\xi$ in a parallel magnetic field $H_\parallel$ one finds 
$\gamma=\frac 16 D  (H_\parallel d)^2$.

\begin{figure}[h!]
\includegraphics[width=1\columnwidth]{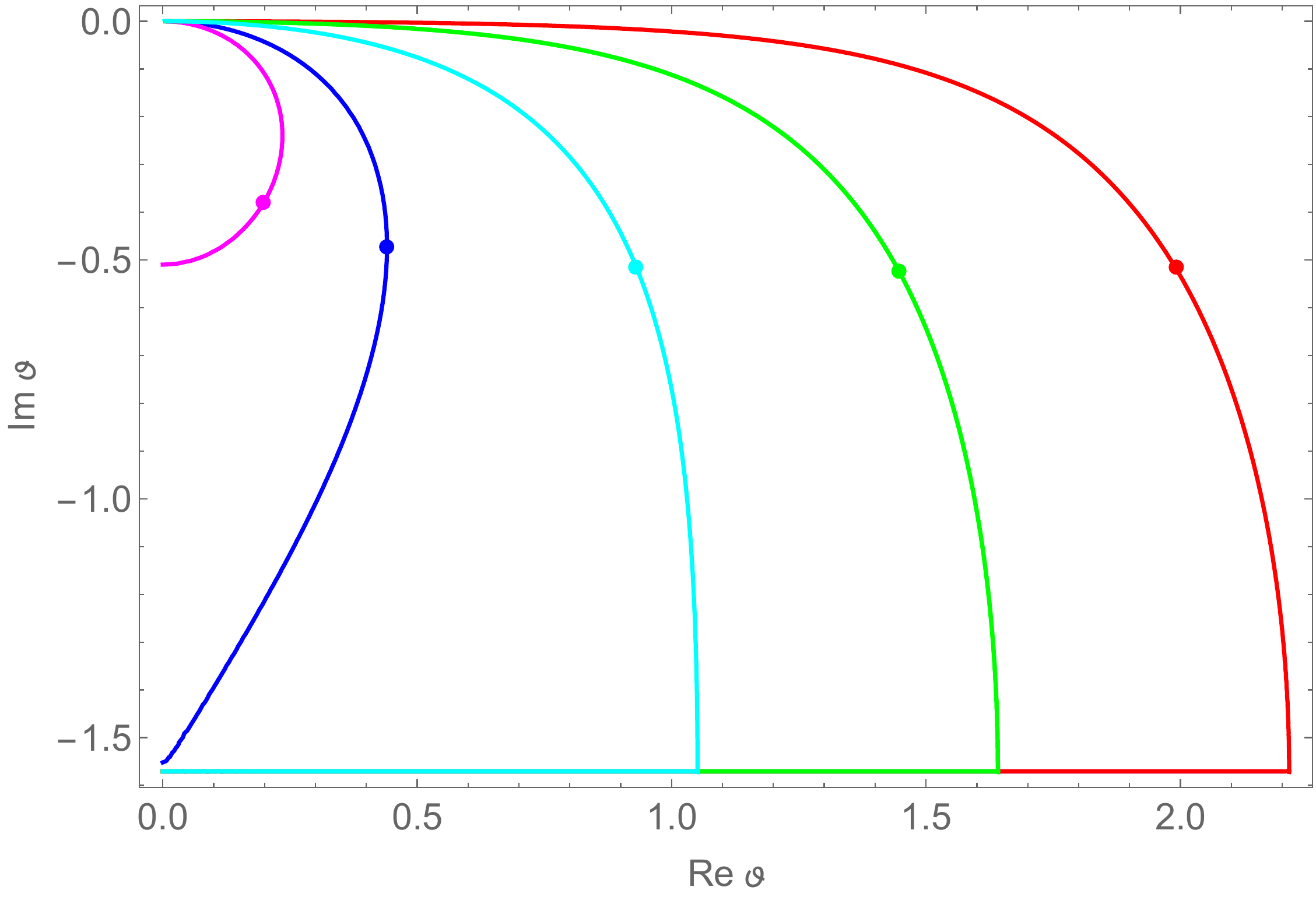}
\caption{ Complex plane of $\vartheta(\e)$ for $\gamma/\Delta_0=  0.01; 0.05;  0.2;  0.456; 0.49$ from right to left. In the gapped case (i.e. $\gamma/\Delta_0<  0.456$)  $\vartheta(0)=-i\pi/2$ and there is a cusp at $\e=\e_g$; eventually  $\vartheta(\infty)\to 0$. Full dots indicate $\epsilon=\Delta$, notice $\mathrm{Im} \vartheta(\Delta)\to -i\pi/6$ as $\gamma\to 0$.  
 }\label{fig:theta}
\end{figure}

Taking retarded and advanced components of the Usadel equation (\ref{SuperCond-Usadel}) without gradients (or equivalently, 
substituting the saddle point ansatz (\ref{SuperCond-classical-ansatz}) -- (\ref{eq:angular-A}) into the action (\ref{eq:S0-no-gradients}) and taking variation over the complex Nambu angles $\vartheta$, $\bar\vartheta$), one finds the saddle point condition: 
\be
\label{eq:Usadel-angular}
\epsilon  =  \Delta \coth{\vartheta}  - i\gamma \cosh{\vartheta}= \Delta \coth{\bar\vartheta}  + i\gamma \cosh{\bar\vartheta}.
\ee   
It's solution is depicted in Fig.~\ref{fig:theta} and admits an important symmetry relation:
\begin{equation}
				\label{eq:theta-symmetry}
\vartheta(-\epsilon) = -\bar\vartheta(\epsilon)\,.
\end{equation}
%

\begin{figure}[h!]
\includegraphics[width=1\columnwidth]{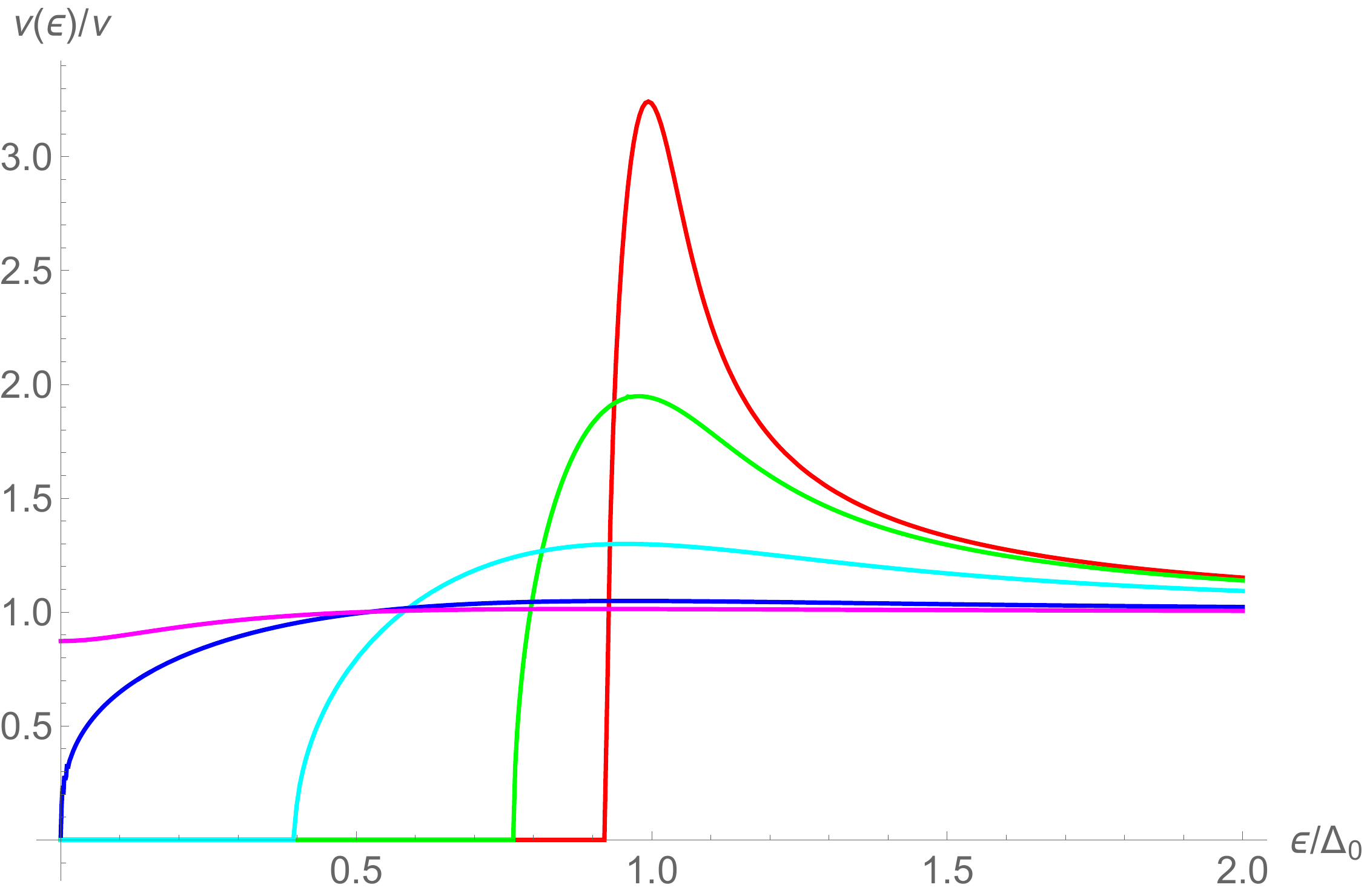}
\caption{ DOS as a function of energy for the same values of TRS breaking parameter $\gamma$ as in Fig~\ref{fig:theta}. 
 }\label{fig:DOS}
\end{figure}

The local DOS is expressed through the Nambu angle as  
\begin{equation}
\label{eq:DOS}
\nu(\epsilon)= {\nu\over 4} \,\mathrm{tr}\big\{\hat \tau^3 \hat \Lambda^R - \hat \Lambda^A \hat\tau^3\big\} = \frac \nu 2  \left[ \cosh\vartheta(\epsilon) + \cosh\bar \vartheta(\epsilon) \right]; 
\end{equation}
it is shown in Fig.~\ref{fig:DOS}.
Within  the energy gap, $|\epsilon| <\epsilon_g$, DOS is zero, i.e.  $\mathrm{Re} \left[ \cosh\vartheta \right]=0$ and thus the angle is  $\vartheta = -i\pi/2+\theta$, with real $\theta$. This brings $ \epsilon  =  \Delta \tanh \theta  - \gamma \sinh \theta$. The right hand side of the latter condition reaches maximum at $\cosh \theta = (\Delta/\gamma)^{1/3}$. Substituting this back into 
Eq.~(\ref{eq:Usadel-angular}) one finds for the energy gap \cite{Larkin1965}
\begin{equation}
			\label{eq:gap}
\epsilon_g = \left(\Delta^{2/3}-\gamma^{2/3}\right)^{3/2}\approx \Delta\left(1-\frac{3}{2}\left(\frac{\gamma}{\Delta}\right)^{2/3}\right),
\end{equation}
where the last approximate relation holds for $\gamma\ll \Delta$.  The gap closes at $\gamma=\Delta$. 
Immediately above the gap, $\epsilon\gtrsim \epsilon_g$, DOS takes the form: 
\be
\label{eq:sqrtDOS}
\nu(\epsilon)= \nu\, \sqrt{\frac 23}\,  \left(\frac{\Delta}{\gamma}\right)^{2/3}\sqrt{\frac{\epsilon-\epsilon_g}{\Delta}}\,.
\ee
At $\epsilon\approx\Delta$ it reaches its maximum value $\nu(\Delta)\approx \frac{\sqrt{3}}{ 4} \nu(4\Delta/\gamma)^{1/3}$ and merges with the 
BCS result $\nu(\epsilon)=\nu\epsilon/\sqrt{\epsilon^2-\Delta^2}$ at $\epsilon-\Delta\propto \Delta^{1/3}\gamma^{2/3}$, see Fig.~\ref{fig:DOS}. 

\begin{figure}[h!]
\includegraphics[width=1\columnwidth]{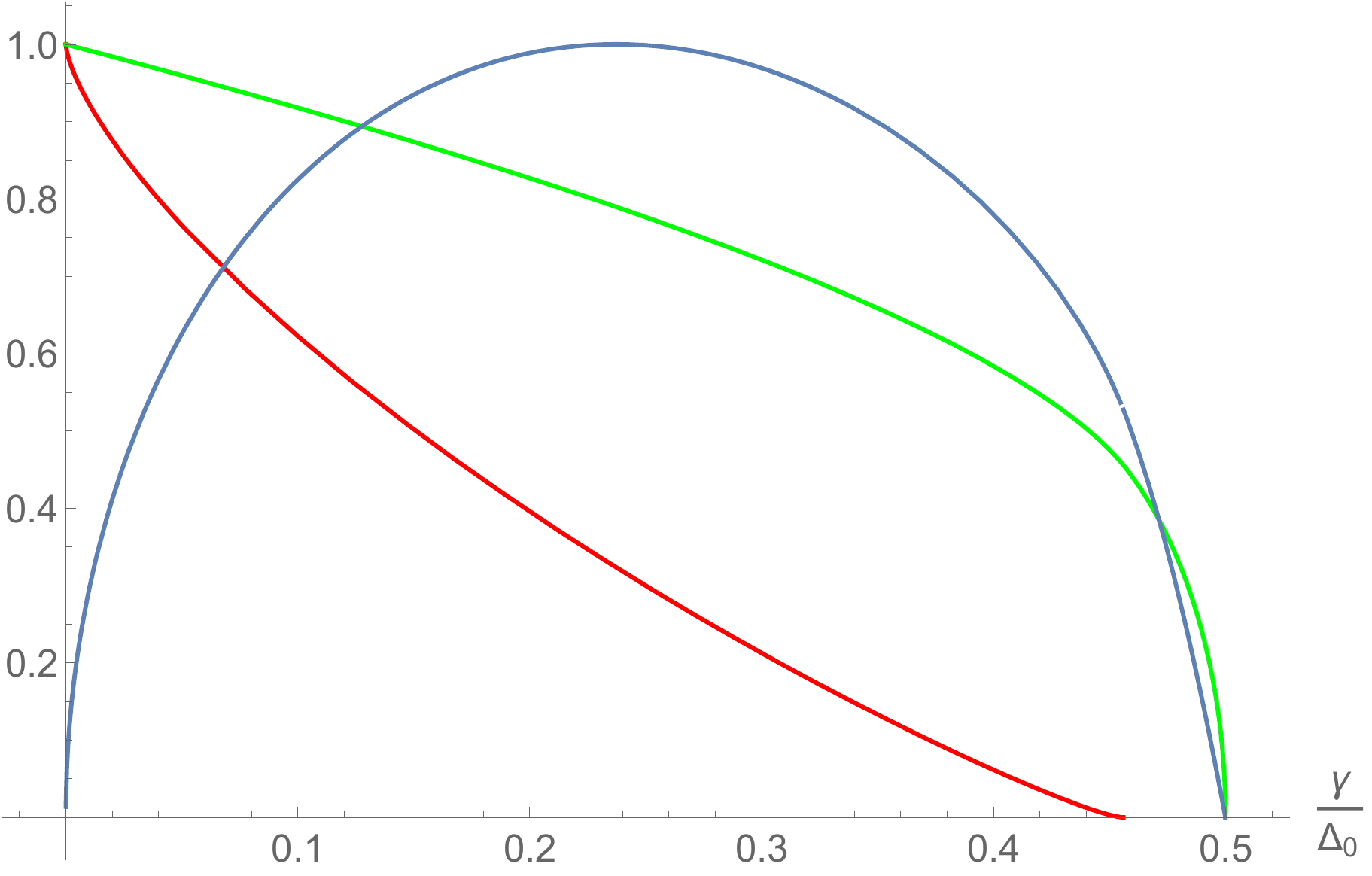}
\caption{Order parameter $\Delta/\Delta_0$ (green) and gap energy $\epsilon_g/\Delta_0$ (red) as 
functions of TRS breaking parameter $\gamma/\Delta_0$. 
The blue line is the super-current density $j_s(\gamma)$  (in arbitrary units), given by Eq.~(\ref{eq:current-2}).   
 }\label{fig:gapless}
\end{figure}

At $T=0$ the self-consistency relation (\ref{eq:self-consistency}) takes the form: 
\begin{equation}
                                                                                     \label{eq:self-consistency-1} 
\Delta =\frac{\lambda}{2}\, \mathrm{Re} \!\! \int\limits_{0}^{\omega_D}\!  d\epsilon \sinh\vartheta = 
\frac{\lambda}{2}\, \mathrm{Re}\!\!  \int\! d\vartheta \,\frac{d\epsilon}{d\vartheta} \, \sinh\vartheta,
\end{equation}
where according to Eq.~(\ref{eq:Usadel-angular}) $ d\epsilon/d\vartheta =-\Delta\sinh^{-2}\vartheta-i\gamma\sinh\vartheta$ and the last integral runs along the contour depicted in Fig.~\ref{fig:theta}. Performing the elementary integration one finds \cite{Larkin1965}
\begin{eqnarray}
                                                                                     \label{eq:self-consistency-2} 
&& \ln\frac{\Delta_0}{\Delta} =\left\{ \begin{array}{ll} 
\pi \gamma/( 4\Delta); & \gamma\leq \Delta, \\
g(\gamma/\Delta); & \gamma >\Delta, 
\end{array}\right. \\
&& g(x)=\ln (x+\sqrt{x^2-1})-\frac{1}{ 2x}  \sqrt{x^2-1} + \frac x 2 \arcsin x^{-1}, \nonumber 
\end{eqnarray}
where $\Delta_0$ is the order parameter at $\gamma =0$. Since $g(x)\to \ln(2x)$ at $x\to \infty$, the self-consistency condition looses a non-trivial solution at $\gamma\geq \Delta_0/2$. On the other hand, the gap closes at $\gamma =\Delta =e^{-\pi/4}\Delta_0\approx 0.456\Delta_0$. Therefore in the narrow range $ 0.456<\gamma/\Delta_0 <0.5 $ the order parameter is finite, while where is no gap in DOS, Fig.~\ref{fig:gapless}. This is the phenomenon of gapless superconductivity. 

For $\gamma\ll \Delta$ one finds from Eq.~(\ref{eq:self-consistency-2}) $\Delta\approx \Delta_0 -\frac \pi 4 \gamma$. The linear in $\gamma$ suppression of the order parameter may be also found from Ginzburg-Landau equation for $T\lesssim T_c$. Notice that 
this  suppression of the order parameter is parametrically weaker than suppression of the gap, Eq.~(\ref{eq:gap}). Therefore for 
the weak breaking of TRS,  $\gamma\ll \Delta_0$, one may drop the distinction between $\Delta$ and $\Delta_0$.

\section{Kinetics of quasiparticles}
\label{sec:kinetics} 

\subsection{Kinetic equation}

The kinetic equations are given by the $(1,2)$ Keldysh  component of the Usadel equation (\ref{SuperCond-Usadel}).
Employing Wigner representation and projecting  onto $\hat \tau^0$ and $\hat\tau^3$ Nambu components, one obtains equations for the longitudinal $F_{\e}^L (\mathbf{r},t)=-F_{-\e}^L (\mathbf{r},t)$ and the transversal $F_{\e}^T(\mathbf{r},t)=F_{-\e}^T (\mathbf{r},t)$  distribution functions:  
\begin{eqnarray} 
				\label{eq:kinetic-L}
&&\!\!\!\!\!\!\!\!  \frac{\nu(\epsilon)}{\nu}\, \partial_t F_{\e}^L \!- \!\nabla_\mathbf{r}\left[D^L(\epsilon)\nabla_\mathbf{r} F_\e^L \right]\! =\!-2I_\mathrm{coll}^L, \\
&&\!\!\!\!\!\!\!\!   \frac{\nu(\epsilon)}{\nu}\, \partial_t F_{\e}^T\!- \! \nabla_\mathbf{r}\left[D^T(\epsilon)\nabla_\mathbf{r} F_\e^T \right] \!+\!  M^T(\epsilon) F_\e^T \!=\! -2 I_\mathrm{coll}^T,
									\label{eq:kinetic-T}
\end{eqnarray}
where local DOS is given by Eq.~(\ref{eq:DOS}) and other parameters are defined as, \cite{LarkinOvchinnikov,BelzigWilhelm,Kamenev2011}:   
\begin{equation}
				\label{eq:DL}
D^{L}(\epsilon)=\frac{D}{4}\,
\mathrm{tr}\left\{\hat\tau^0 - \hat{Q}^{R}\hat{Q}^{A}\right\}
=D\cosh^2 \left(\frac{\vartheta-\bar\vartheta}{2}  \right),
\end{equation}
\begin{equation}
				\label{eq:DT}
D^{T}(\epsilon)=\frac{D}{4}\,
\mathrm{tr}\left\{\hat\tau^0 - \hat\tau^3 \hat{Q}^{R}\hat\tau^3\hat{Q}^{A}\right\}
=D\cosh^2 \left(\frac{\vartheta+\bar\vartheta}{2}  \right); 
\end{equation} 
\begin{eqnarray}
				\label{eq:MT}
M^{T}(\epsilon)&=&\frac{1}{2}\,
\mathrm{tr}\left\{\hat{Q}^{R}\hat\Delta + \hat\Delta\hat{Q}^{A}\right\}
=i\Delta\big(\sinh\vartheta - \sinh \bar \vartheta\big) \nonumber \\
&=& 2\gamma \cosh^2 \left(\frac{\vartheta+\bar\vartheta}{2}\right)  |\sinh\vartheta|^2.
\end{eqnarray}
The mass, $M^{T}(\epsilon)$, exists only in the absence of TRS. It goes to zero at large energy as $M^T\to 2\gamma\Delta^2/\epsilon^2$, but acquires a large value  $M^T(\epsilon_g)=2\Delta^{4/3}\gamma^{-1/3}$ near the gap. 
Such a mass provides a rapid decay of  the transversal component of the distribution function to zero.  We thus focus here only on the slow longitudinal relaxation. 

\begin{figure}[h!]
\includegraphics[width=1\columnwidth]{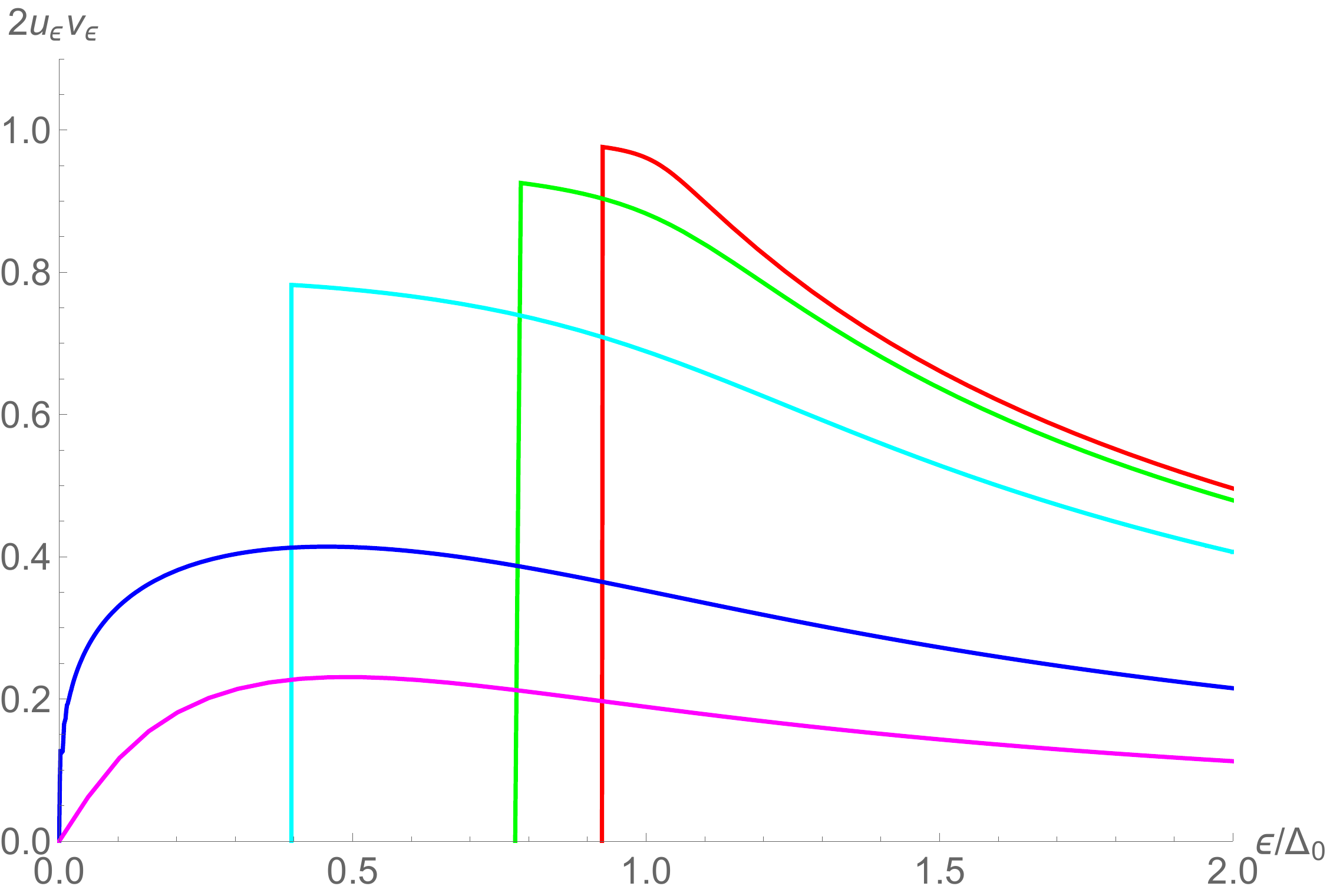}
\caption{$2u_\e v_\e$ as a function of energy for the same values of TRS breaking parameter $\gamma$ as in Fig~\ref{fig:theta}. 
 }\label{fig:uv}
\end{figure}

The corresponding collision integral is given by Eq.~(\ref{eq:collisions-normal})  (with factor $1/2$ to compensate for Nambu doubling of the degrees of freedom), where one should use the Nambu-rotated $\Lambda_\e$-matrices, Eq.~(\ref{eq:Utau3U}). This yields, e.g.: 
\begin{eqnarray}
&&\mathrm{tr_N}\left\{\hat\tau^0 \left[\hat\gamma^{cl}  \hat\Lambda_{\epsilon'} \hat\gamma^{cl}  \hat\Lambda_{\epsilon} - \hat\Lambda_{\epsilon} \hat\gamma^{cl}  \hat\Lambda_{\epsilon'}\hat\gamma^{cl} 
\right]^{(1,2)}_K\right\}  \\ && = 2(F_\e^L-F_{\e'}^L)\,\mathrm{Re}\left[\cosh(\vartheta-\vartheta') + \cosh(\vartheta-\bar\vartheta')\right], \nonumber 
\end{eqnarray}
where $\vartheta=\vartheta(\e)$ and $\vartheta'=\vartheta(\e')$.
As a result, the kinetic equation for the quasiparticles occupation number $f_\e=(1-F^L_\e)/2$ acquires a form:
\begin{eqnarray} 
				\label{eq:kinetic-f}
&&\frac{\nu(\epsilon)}{\nu}\, \partial_t f_{\e} \!- \!\nabla_\mathbf{r}\left[D^L(\epsilon)\nabla_\mathbf{r} f_\e \right]\! = I_\mathrm{coll}^L[f_\e], \\
&&I_\mathrm{coll}^L[ f_\epsilon]\!=\!\!\int\!\frac{d\e'}{2\pi} M_{\e,\e'}^S
[N_{\omega} f_{\e'}(1-f_{\e}) \!-\! (1+N_{\omega}) f_\e(1-f_{\e'})], \nonumber 
\end{eqnarray}
where superconducting phonon matrix element  is:  
\begin{eqnarray}
				\label{eq:dupe-matrix-element}
M^S_{\e,\e'}&=&\frac 1 2\, \mathrm{Re}\left[\cosh(\vartheta-\vartheta') + \cosh(\vartheta-\bar\vartheta')\right] M^N_{\e-\e'} \nonumber \\
&=&\frac{\nu(\e)}{\nu} \frac{\nu(\e')}{\nu} \left[1 - 4u_\e v_\e u_{\e'} v_{\e'} \right] M^N_{\e-\e'},
\end{eqnarray}
where $\omega=\e-\e'$, the normal state matrix element $M^N_\omega$ is given by Eqs.~(\ref{eq:matrix-element-normal}), (\ref{eq:matrix-element-normal-1})  and DOS $\nu(\e)$ is given by Eqs.~(\ref{eq:DOS}), (\ref{eq:sqrtDOS}).
Motivated by standard TRS notations, we introduced    
\begin{equation}
				\label{eq:2uv}
2u_\e v_\e \equiv  \frac{\mathrm{Re}[\sinh\vartheta]}{\mathrm{Re}[\cosh \vartheta]},
\end{equation}
which is only defined for $|\e|>\e_g$, see Fig.~\ref{fig:uv}.  Employing Eqs.~(\ref{eq:Usadel-angular})--(\ref{eq:sqrtDOS}), one may show that 
\begin{equation}
					\label{eq:2uv-1}
2u_\e v_\e \approx \left\{ \begin{array}{ll} 
\Delta/|\e|; & \quad \e-\Delta \gg \Delta^{1/3}\gamma^{2/3};\\
\sqrt{1-(\gamma/\Delta)^{2/3}}; & \quad |\e-\Delta| \lesssim \Delta^{1/3}\gamma^{2/3}
\end{array} \right. 
\end{equation}
Since $\nu(\e')=0$ for $|\e'|< \e_g$, while $\left[1 - 4u_\e v_\e u_{\e'} v_{\e'} \right] $ tends to a constant,  one finds  $M^S_{\e,\e'<\e_g}=0$: as expected the final energy $\e'$ has to be outside the spectral gap.   For TRS superconductors these results appeared in Ref.~\cite{Reizer}. 

The factor $\nu(\e)/\nu$ on the right hand side of Eq.~(\ref{eq:dupe-matrix-element}) is cancelled against the same on the left hand side of Eq.~(\ref{eq:kinetic-f}). As a result, one finds for the ``out''  electron-phonon relaxation rate: 
\begin{equation}
				\label{eq:out}
\frac{1}{\tau_{e:ph}}=\!\!\!\int\limits_{|\e'|>\e_g} \!\!\!\! \frac{d\e' \nu(\epsilon')}{2\pi\nu}(1+N_{\omega})(1-f_{\e'})\left[1 - 4u_\e v_\e u_{\e'} v_{\e'} \right] M^N_\omega. 
\end{equation}
The energy integral here may be further subdivided onto positive, $\e'>\e_g$, and negative $\e'<-\e_g$ regions, representing the inelastic {\em scattering} and {\em recombination} processes correspondingly:
\begin{equation}
\frac{1}{\tau_{e:ph}(\e)}=\frac{1}{\tau_{e:ph}^\mathrm{sc}(\e)}+\frac{1}{\tau_{e:ph}^\mathrm{rec}(\e)}, 
\end{equation}
where
\begin{eqnarray}
&& \frac{1}{\tau_{e:ph}^\mathrm{sc}}\!=\!\!\!\int\limits_{\e_g}^\infty \!\! \frac{d\e' \nu(\epsilon')}{2\pi\nu}(1\!+\!N_{\e-\e'})(1\!-\!f_{\e'})\!\left[1\! -\! 4u_\e v_\e u_{\e'} v_{\e'} \right] \! M^N_{|\e-\e'|}; \nonumber \\ 
&& \frac{1}{\tau_{e:ph}^\mathrm{rec}}\!=\!\!\!\int\limits_{\e_g}^\infty \!\! \frac{d\e' \nu(\epsilon')}{2\pi\nu}(1+N_{\e+\e'})f_{\e'}\!\left[1\! -\! 4u_\e v_\e u_{\e'} v_{\e'} \right] \! M^N_{|\e+\e'|}.
		\label{eq:out-sc}
\end{eqnarray}  
To obtain recombination time we changed integration variable $\e'\to -\e'$ and used that $F_{-\e'}^L=-F^L_{\e'}$ and therefore 
$(1-f_{-\e'})=f_{\e'}$. We also employed Eq.~(\ref{eq:theta-symmetry}), which insures that both $\nu(\e')$ and $2u_{\e'}v_{\e'}$ are even functions. For low concentration of non-equilibrium quasiparticles $f_{\e'}\ll 1$ the recombination processes may be disregarded, even though their matrix element tends to be larger.

\section{Kinetics of quasiparticles trapping}
\label{sec:trapping}

\subsection{Trapping rate} 

We now focus on trapping of non-equilibrium quasiparticles within the regions with the locally suppressed energy gap. 
Such suppression is often achieved by breaking TRS, resulting in a spatially  dependent TRS breaking parameter $\gamma(\mathbf{r})$. For example, an isolated Abrikosov vortex brings $\gamma(r) = \frac12\Delta(\xi/r)^2$.

Quasiparticles with an initial energy $\e >\Delta$ diffuse to the regions with the suppressed gap $\e_g(\mathbf{r})<\Delta$. There they can inelastically scatter  to a final energy $\e'$ within the window $\e_g(\mathbf{r}) < \e' <\Delta$ by emitting an acoustic phonon. As a result, they end up being trapped within the spatial region $\e_g(\mathbf{r}) <\e'$, due to Andreev reflections from its boundaries.  We evaluate the corresponding trapping rate, assuming very low phonon temperature, $N_\omega\ll 1$, and small concentration of non-equilibrium quasiparticles, i.e. $f_{\e'}\ll 1$. As a result the trapping rate is found as: 
\begin{equation}
				\label{eq:trapping}    
\frac{1}{\tau_\mathrm{tr}}= \frac{b_j\pi\Omega_d}{(2\pi)^d}\,  \frac{Dp_F^2 }{\rho_m v_j^{d+2}}\!\int\limits_{\e_g}^\Delta \! \frac{d\e' \nu(\epsilon')}{2\pi\nu} \left[1 - 4u_\e v_\e u_{\e'} v_{\e'} \right] (\epsilon\! -\! \epsilon')^d\!, 
\end{equation}
where the coefficients $b_j$ are defined after Eq.~(\ref{eq:matrix-element-normal-1}).

There are two distinct limits for the trapping rate distinguished by the comparison of the relative excess energy of  nonequilibrium quasiparticles,  $\delta_\e\equiv (\e-\Delta)/\Delta$, and the relative energy range affected by breaking of TRS, $(\gamma/\Delta)^{2/3}$. Carrying out the integration in Eq.~(\ref{eq:trapping}) with the help of Eqs.~(\ref{eq:sqrtDOS}), (\ref{eq:2uv-1}), one finds;
\begin{equation}
				\label{eq:traping-rate} 
\frac{1}{\tau_\mathrm{tr}}\propto \frac{Dp_F^2\Delta^{d+1}}{\rho_m v_j^{d+2}}\left(\frac{\gamma}{\Delta}\right)^{\!\!\frac13} 
\! \left\{ \begin{array}{lc}
\!\!\!(\gamma/\Delta)^{\frac{2}{3}(d+1)}; & \delta_\e < (\gamma/\Delta)^{\frac23},\\
\!\!\!\delta_\e^{d+1} ;   & \hskip -.6cm   (\gamma/\Delta)^{\frac23} < \delta_\e\lesssim 1. 
\end{array} \right. 
\end{equation}

In most metals the longitudinal sound velocity is about twice that of the transversal one. As a result, the {\em transversal} phonons
are about an order of magnitude more efficient in trapping the non-equilibrium quasiparticles than the longitudinal ones. Hereafter we thus restrict ourselves exclusively to the transversal waves. Notice that the transversal phonons are coupled to electrons due to impurity displacement mechanism, which is only present in the disordered limit $ql=\omega l/v_t\lesssim 1$. 
The characteristic length scale $v_t/\omega$ is typically in the range 10-100 nm. We shall assume that the characteristic thickness of superconducting films is larger than that and  put $d=3$ in the subsequent estimates.

\subsection{Trapping power of a single vortex} 

We now evaluate the total trapping power of a vortex in a superconducting film,  defined as a spatial integral of the local trapping rate (\ref{eq:trapping}), $P=\int d^2\mathbf{r}/\tau_\mathrm{tr}(\mathbf{r})$. This quantity may then be used as a sink term in the macroscopic 2D diffusion equation for the density of non-equilibrium quasiparticles, $n(\mathbf{r},t)$,
\begin{equation}
\partial_t n - \nabla[D\nabla n] = - P \, \delta^{(2)}(\mathbf{r})\, n,
\end{equation}
where the vortex is assumed to be placed at $\mathbf{r}=0$.

For an isolated vortex the TRS breaking parameter is a function of the distance $r$ from the core $\gamma=\frac{1}{2}\Delta (\xi/r)^2$. The trapping rate at small $\gamma$ scales as $\gamma^{1/3}\sim r^{-2/3}$, and thus the  integral in the definition  of the trapping power 
is dominated by large distances from the vortex core.   Employing Eq.~(\ref{eq:trapping}), one finds: 
\begin{equation}
				\label{eq:tracing-power}
P =	\!  \frac{D(p_F\xi)^2\Delta^4 }{10\pi \rho_m v_t^{5}} \!
\left[ \left(\frac{r_c}{2\xi}\right)^{\!\!2} \left(\left(1+\delta_\epsilon\right)^4 -\delta_\epsilon^{4}\right) \! +\! \left(\frac{R}{\xi}\right)^{\!\!4/3}\!\!\!\delta_\epsilon^{4} \right]\!.  			
\end{equation}
The first term in the square brackets here is the contribution of the vortex core, which we model  as a normal cylinder with the radius $r_c$. The second term is coming from the outer periphery of the vortex core with $R$ being it's effective outer radius.  It  is determined by either a distance between vortices, a penetration depth, or the condition that $\Delta-\epsilon_g(R)\approx T$, where $T$ is the phonon temperature. Indeed, beyond such a radius the trap is too shallow and trapping is not effective because of the activation escape. This leads to $(R/\xi)^{4/3} \approx \Delta/T$ and and allows us to rewrite the last term in the brackets of Eq. (\ref{eq:tracing-power}) as $(\delta_\epsilon\Delta^{1/4}/T^{1/4})^4$. The peripheral trapping may dominate, if the typical quasiparticles excess energy grossly exceeds the phonon temperature.

We now use the parameters of devices investigated in experiment \cite{Chen2014} to estimate $P$ with the help of Eq. (61) and compare it with the experimental findings. In Ref.~\cite{Chen2014}  the trapping power of individual vortices in an aluminum film was measured to be $P=6.7\times 10^{-2}$ cm$^2$/s at the base temperature of $T=20$ mK. The relevant parameters of the film were \cite{Chen2014} $D=18$ cm$^2$/s; $v_t=3.0\times 10^5$ cm/s; $\rho_m=2.7$ g/cm$^3$; $E_F=11.7$ eV; $\Delta=1.8\times 10^{-4}$ eV. With these parameters one finds 
$D(p_F\xi)^2\Delta^4/(\rho_m v_t^{5})=6.8\times 10^{-3}$ cm$^2$/s. As a result, the core contribution to the trapping power in Eq.~(\ref{eq:tracing-power}) is about two orders of magnitude smaller than the observed value for a reasonable estimate of $r_c$ and $\delta_\e$. 
At the base temperature, $(R/\xi)^{4/3}\approx \Delta/T=10^2$, the peripheral contribution may provide trapping of the right order of magnitude only if $\delta_\epsilon\sim 1$. For the geometry of devices in Ref.~\cite{Chen2014}, there are no reasons to expect that the quasiparticles are so "hot" in the vicinity of the vortices. Therefore, albeit the peripheral contribution adds to the trapping power, it is not sufficient to explain the observed value of $P$.

\subsection{Trapping by a current-carrying constriction}

Trapping rate was also measured \cite{Siddiqi2014} in a nano-bridge closed by a flux-biased superconducting loop. The flux bias was creating a super-current flowing through the constriction. The super-current breaks TRS and thus suppresses the the energy gap in the nano-bridge itself as well as 
in the adjacent leads, carrying the stray currents. Assuming $3D$ leads, the stray current density may be estimated as $j_\mathrm{s}(r)=I_\mathrm{s}/(2\pi r^2)$,  where $r$ is distance from the constriction and $I_\mathrm{s}$ is the total super-current through the constriction. 

To apply our theory we need to find a relation  between the local supper-current density $j_\mathrm{s}(r)$ and the local 
TRS breaking parameter $\gamma(r)$.   Assuming an applied vector potential $\mathbf{A}(r)$, the super-current density  is obtained by  variation of the action (\ref{eq:S0-supercond}) over the (quantum component of the) vector potential and is given by
\begin{equation}
				\label{eq:current}
-j_s\!=\!A{e\nu D} \mathrm{Im}\!\! \int\limits_0^\infty\!\! d\epsilon\, F^L_\epsilon  \sinh^2\vartheta\! =\!A{e\nu D}\mathrm{Im}\!\!\!\! \int\limits_{\vartheta(0)}^0\!\! \!\!d\vartheta \frac{d\epsilon}{d\vartheta}  \sinh^2\!\vartheta,
\end{equation}
where in the second equation we put $T=0$ and changed the integration variable to $\vartheta$. 
We will also assume that the vector potential $A$, which creates the super-current is the sole source of the breaking TRS symmetry (i.e. no additional magnetic field is present) and thus $\gamma =2DA^2$. Performing the integration one finds (we traded here $A$ for $\gamma$):
\begin{eqnarray}
				\label{eq:current-2}
j_s &=& e\nu \gamma^{3/2} \sqrt{\frac { D}2}\,  \mathrm{Re}\, h\left(\frac\Delta\gamma\right); \\
h(x) &=& x \sin^{-1}x- \frac 23 \left (1- \sqrt{1-x^2}\left(1+\frac 12 x^2\right)\right). \nonumber
\end{eqnarray}
Notice that $\Delta=\Delta(\gamma)$ according to Eq.~(\ref{eq:self-consistency-2}). The resulting critical current $j_s(\gamma)$ is plotted in Fig.~\ref{fig:gapless}.  For small $j_s$  the effective TRS breaking parameter $\gamma$ is found from Eq.~(\ref{eq:current-2}) as  
\begin{equation}
				\label{eq:gamma-current}
\gamma(r)= \frac{8}{\pi^2} \Delta \left(\frac{j_s(r)}{e\nu\xi\Delta^2 }\right)^2= \frac{2\Delta}{\pi^4}  \left(\frac{I_s}{e\nu\xi^3\Delta^2 }\right)^2 \left(\frac{\xi}{r }\right)^4.
\end{equation}

According to Eq.~(\ref{eq:traping-rate}) the trapping rate far from the constriction  (i.e. for small $\gamma$)  scales as $\tau_\mathrm{tr}^{-1}(r)\sim \gamma^{1/3}\sim I_s^{2/3} r^{-4/3}$. To calculate the total trapping rate $\tau_T^{-1}$ of the constriction, measured in Ref.~[\onlinecite{Siddiqi2014}], one integrates this expression over the volume of the leads and multiplies by 
$n_{qp}$ -- concentration of non-equilibrium quasiparticles (in notations of Ref.~[\onlinecite{Siddiqi2014}] $n_{qp}=x_{qp}\nu\Delta$, where $x_{qp}$ is dimensionless fraction of broken Cooper pairs). The aforementioned volume integral is  coming from large distances $R$ and thus  $  \tau_T^{-1}\sim  I_s^{2/3} R^{5/3}$. As explained above,
the outer radius $R$ is limited by thermally activated escape and is estimated from $(\gamma(R)/\Delta)^{2/3} =T/\Delta$. This leads to   $R\sim I_s^{1/2} T^{-3/8}$.
As a result, the trapping rate of the constriction, coming from its outer periphery, is given by: 
\begin{equation}
			\label{eq:constriction}
\frac{1}{\tau_T} \propto \frac{Dp_F^2(\Delta\delta_\e)^4 }{ \rho_m v_t^{5}} n_{qp}\xi^3 \left(\frac{I_s}{e\nu\xi^3\Delta^2 }\right)^{2/3}  \left(\frac{\Delta}{T}\right)^{5/8} .
\end{equation}
(For 2D pattern of stray currents one finds $\tau_T^{-1}\sim I_s^2 T^{-1}$.) Both  dependencies on current and temperature are in a qualitative agreement with the data of Ref.~[\onlinecite{Siddiqi2014}]. The 
``core'' contribution, due to trapping on localized Andreev bound states, was calculated in   Ref.~[\onlinecite{Siddiqi2014}] and found to be about two orders of magnitude less than the observed value. 
The peripheral trapping, discussed here, may well account for this discrepancy, though quantitative comparison is impeded by the uncertainty in $\delta_\e$ and $n_{qp}$.

\section{Discussion of the results}
\label{sec:discussion}

We have developed a unified theory for treating electron-phonon kinetics in disordered normal metals and 
superconductors, including superconductors with broken TRS. The latter case is particularly important for evaluation of the 
trapping rate of non-equilibrium quasiparticles in the regions, where the energy gap is suppressed by a local magnetic field or a super-current.  Quasiparticle traps are proven to be useful for increasing coherence time of superconducting qubits. 
 
Our theory shows that the trapping rate, $\tau_\mathrm{tr}^{-1}$, is a very sensitive function of the TRS breaking parameter
 $\gamma$, which  at low temperature scales as  $\tau_\mathrm{tr}^{-1}\propto \gamma^{1/3}$. As a result, even the regions with a weak breaking of TRS, such as a far periphery of a vortex or a constriction, may provide a significant contribution to the overall trapping power of ``hot'' quasiparticles.  The quantitative comparison with the experiment [\onlinecite{Siddiqi2014}] requires detailed knowledge of non-equilibrium quasiparticles energy distribution, which is not available at the moment. Our estimates show that in order to account for the observed trapping rates, the non-equilibrium quasiparticles excess energy should be $\e-\Delta \sim \Delta\gg T$, where $T$ is the phonon bath temperature.

\section{Acknowledgments} We are grateful to M. Feigelman, A. Shtyk and V. Yudson for illuminating discussions.
This work was supported by the DOE contract DEFG02-08ER46482.

\appendix

\section{Alternative derivation of the electron-phonon action} 
\label{sec:alternative} 

Here we provide an alternative derivation of Eqs.~(\ref{eq:S-u}), which is based on first principles Coulomb interactions between the electrons and the lattice as well as impurity drag by the lattice displacements. The first of these effects leads to to the standard Coulomb action
\begin{equation}
\label{eq:Coulomb-action}
S_C\!=\!\!\!\int\!\! dt\! \left[\frac 1 2 \sum\limits_\mathbf{q} \varphi_{\mathbf{q},t} U_C^{-1} \varphi_{-\mathbf{q},t} +\!
\sum\limits_\mathbf{r} \varphi_{\mathbf{r},t}(\rho_0\mathrm{div}\,\mathbf{u}-\rho_e)_{\mathbf{r},t}\!\right]\!, 
\end{equation}
where $\varphi_{\mathbf{r},t}$ is the fluctuating scalar potential, $U_C=4\pi e^2/q^2$ is the bare Coulomb interaction,  
$\rho_e(\mathbf{r},t) = \bar\psi(\mathbf{r},t) \psi(\mathbf{r},t) -\rho_0$ is the excess electron density, while $\rho_0\mathrm{div}\,\mathbf{u}$ is the excess lattice density.  

The second effect is more subtle and pertains to the
disordered limit $ql\ll 1$, where $l = v_F \tau $ is the elastic
mean free path and $\tau$ is the elastic mean free time. It
originates from the fact that the impurities are frozen
into the crystal lattice and therefore are also subject to
the displacement $\mathbf{u}(\mathbf{r},t)$ \cite{Pippard,Tsuneto,Schmid,
Reizer,Yudson,Shtyk}. Therefore hitherto static
random disorder potential becomes a dynamic object
$V_\mathrm{dis}(\mathbf{r})\rightarrow  V_\mathrm{dis}(\mathbf{r}+\mathbf{u}(\mathbf{r},t))$. 
It is convenient to shift $\mathbf{r}$ to write the interaction of the electron
density with the disorder potential as
\begin{equation}
H_\mathrm{dis}=\sum\limits_\mathbf{r}V_\mathrm{dis}(\mathbf{r}) \rho_e(\mathbf{r}-\mathbf{u}(\mathbf{r},t),t)
\end{equation}
Performing averaging over the Gaussian distribution of short-ranged disorder, one finds the following action
\begin{equation}
iS_\mathrm{dis} = -\frac{1}{4\pi\nu\tau}\! \int\!\!\!\int\! dt dt' \sum\limits_\mathbf{r} 
\bar\psi_{\mathbf{r}-\mathbf{u},t} \psi_{\mathbf{r}-\mathbf{u},t}  
\bar\psi_{\mathbf{r}-\mathbf{u}',t'} \psi_{\mathbf{r}-\mathbf{u}',t'},  
\end{equation}
where $\mathbf{u}=\mathbf{u}(\mathbf{r},t)$ and $\mathbf{u}'=\mathbf{u}(\mathbf{r},t')$. One can now rearrange the fermionic fields and decouple the four-fermion action with the help of the non-local in time field $Q_{t,t'} (\mathbf{r})$. This leads to the following term:
\begin{equation}
					\label{eq:QL}
\bar\psi_{\mathbf{r}-\mathbf{u},t} Q_{t,t'} (\mathbf{r}) \psi_{\mathbf{r}-\mathbf{u}',t'} \approx 
\bar\psi_{\mathbf{r},t}\left[ Q_{t,t'} (\mathbf{r}) -\hat{\cal L}_1 + \frac 1 2 \hat{\cal L}_2\right]  \psi_{\mathbf{r},t'}. 
\end{equation}
We have expanded fermionic fields to the second order in
the displacement $\mathbf{u}$, which brings the two operators:
\begin{eqnarray}
\hat{\cal L}_1 &=& \stackrel{\leftarrow}{\nabla} \cdot \mathbf{u}\, Q_{t,t'} (\mathbf{r})  + 
Q_{t,t'} (\mathbf{r})  \mathbf{u}'\cdot \stackrel{\rightarrow}{\nabla};\\
\hat{\cal L}_2 &=& \stackrel{\leftarrow}{\nabla} \stackrel{\leftarrow}{\nabla} \cdot\!\cdot \mathbf{u}  \mathbf{u}\, Q + 
\stackrel{\leftarrow}{\nabla} \cdot \mathbf{u} \, 2 Q \mathbf{u}'\cdot \stackrel{\rightarrow}{\nabla} + 
Q\, \mathbf{u}'\mathbf{u}'\cdot \!\cdot \stackrel{\rightarrow}{\nabla} \stackrel{\rightarrow}{\nabla},\nonumber
\end{eqnarray}
where the arrows above the gradient operators show direction of the differentiation in the context of Eq. (\ref{eq:QL}).
The action is now quadratic in the unshifted fermionic fields which may be integrated out in the standard way, leading to the determinant:
\begin{equation}
\label{eq:trace-log}
\mathrm{Tr} \log \left\{ G_0^{-1}- \varphi +\frac{i}{2\tau} \left[ Q-\hat{\cal L}_1 + \frac 1 2 \hat{\cal L}_2\right]\right\}, 
\end{equation}
where $\varphi = \varphi_{\mathbf{r},t}$ is the scalar potential coming from the
Coulomb interactions, Eq. (\ref{eq:Coulomb-action}). 

From this point on, one proceeds along the standard root
of deriving Keldysh non-linear sigma-model \cite{Kamenev2011}. 
To this end one passes to the Keldysh $2 \times 2$ structure, by splitting the contour on forward and backward branches and performing the Keldysh rotation. 
One then realizes that the soft diffusive modes of the action are described by the manifold $\hat Q^2 = 1$ and therefore one can write $\hat Q = \hat {\cal R}^{-1}\hat \Lambda \hat{\cal R}$, where $\hat \Lambda$ is the Green function in coinciding spatial points,
Eq.~(\ref{eq:Lambda}). This way Eq.~(\ref{eq:trace-log}) may be rewritten as:
\begin{equation}
\label{eq:trace-log-1}
\mathrm{Tr} \log\! \left\{\! 1 \!+\! \hat G \hat{\cal R} [G_0^{-1},\hat{\cal R}^{-1}] \!-\!  \hat G \hat{\cal R}\! \left[  \hat \varphi\! +\! \frac{i}{2\tau} \hat{\cal L}_1\! -\! \frac{i}{4\tau} \hat{\cal L}_2\right]\!\hat{\cal R}^{-1}\!\right\}\!.
\end{equation}
Finally, one expands the logarithm here to the lowest non-vanishing orders. This way one obtains the standard non-linear sigma-model action (first neglecting $\hat{\cal L}_{1,2}$ terms):
\begin{equation}
\label{eq:S0-again}
iS_0=\frac{i\nu}{2} \mathrm{Tr} \{ \hat \varphi\hat \sigma^1  \hat \varphi\} - \frac{\pi\nu}{4}\mathrm{Tr} \{ D(\partial_\mathbf{r} \hat Q)^2 -4\partial_t \hat Q-4i\hat\varphi\hat Q\}. 
\end{equation}
The first term on the right hand side here represents static polarizability (i.e. screening) of the electronic band. It comes from the so-called retarded-retarded and advanced-advanced loops. The dynamic screening is encoded in $\pi\nu \mathrm{Tr}\{\hat\varphi\hat Q\}$  term along with fluctuations of the $\hat Q$ field around itÕs stationary point $\hat\Lambda$.

We focus now onto the phonon-induced $\hat{\cal L}_{1,2}$ terms, which originate from the motion of the impurities relative to the electronic liquid. It is easy to see that the first order in $\hat{\cal L}_{1}$ vanishes. One is thus left with the three terms: (i) first order in $\hat{\cal L}_{1}$ and in $i\hat{\cal R} \mathbf{v}_F\cdot \stackrel{\rightarrow}{\nabla}\hat{\cal R}^{-1}$; (ii) first order in $\hat{\cal L}_{2}$ and (iii)
second oder in $\hat{\cal L}_{1}$. A straightforward, but somewhat lengthy evaluation of these three terms results it 
\begin{eqnarray}
&&\!\!\!\!\! iS_\mathrm{(i)}= -i\pi\nu\frac{v_Fp_F}{d} \mathrm{Tr} \{ \hat{\mathbf{u}}\cdot \nabla \hat Q \} =i\pi\rho_0  \mathrm{Tr}\{ \mathrm{div} \hat{\mathbf{u}} \, \hat Q\}; \label{eq:Si}  \\
&&\!\!\!\!\!  iS_\mathrm{(ii)}= -i\frac{\pi\nu}{2\tau} \frac{p_F^2}{d} \, \mathrm{Tr} \{ \hat{\mathbf{u}}\cdot  \hat{\mathbf{u}} -  \hat{\mathbf{u}}\, \hat Q \hat{\mathbf{u}}\, \hat Q \}; \label{eq:Sii} \\
&&\!\!\!\!\!  iS_\mathrm{(iii)}=  i\frac{\pi\nu}{2\tau} \frac{p_F^2}{d} \, \mathrm{Tr} \{ \hat{\mathbf{u}}\cdot  \hat{\mathbf{u}} -  \hat{\mathbf{u}}\, \hat Q \hat{\mathbf{u}}\, \hat Q \}    \label{eq:Siii} \\ 
&&\quad \quad\quad +\, \frac{\pi\nu D\, p_F^2}{4}\,\,  \mathrm{Tr}\big\{ [\hat Q\,,\partial^\mu \hat{\mathbf{u}}^\nu]  [\hat Q\,,\partial^\eta \hat{\mathbf{u}}^\lambda] \big\}\, \Upsilon_{\mu\nu,\eta\lambda}\nonumber ,
\end{eqnarray}
where $\Upsilon_{\mu\nu,\eta\lambda}$ is given by Eq.~(\ref{eq:upsilon}). 
Notice that the leading orders in $S_\mathrm{(ii)}$ and $S_\mathrm{(iii)}$ exactly cancel each other. The second sub-leading term in Eq.~(\ref{eq:Siii}) originates from gradient operators in $\hat{\cal L}_{1}$ acting on displacements $\hat{\mathbf{u}}$, as opposed to the Green functions $G$. 

The scalar linear coupling $S_\mathrm{(i)}$ may be combined with the  potential term in Eq.~(\ref{eq:S0-again}) by shifting the potential  $\hat\varphi\to \hat\phi=\hat\varphi +\frac{\rho_0}{\nu} \mathrm{div} \hat{\mathbf{u}}$. In the limit 
of the strong Coulomb interactions, $U_C\to \infty$ in Eq.~(\ref{eq:Coulomb-action}), this allows to  eliminate 
Fr\"ohlich deformation potential electron-phonon coupling. Indeed, the static screening  $\frac{\nu}{2} \mathrm{Tr} \{ \hat \varphi\hat \sigma^1  \hat\varphi\}$ in  Eq.~(\ref{eq:S0-again}) along with the interaction term $\hat\varphi\hat\sigma^1\rho_0\mathrm{div}\,\hat{\mathbf{u}}$ in Eq.~(\ref{eq:Coulomb-action}) upon the aforementioned shift results in $\frac{\nu}{2} \mathrm{Tr} \{ \hat \phi\hat \sigma^1  \hat \phi\}-\frac{\rho_0^2}{2\nu} \mathrm{div} \hat{\mathbf{u}}\, \hat\sigma^1 \mathrm{div} \hat{\mathbf{u}}$.
The first term here stay for the screened electron-electron interactions, unaffected by lattice displacement, while the second one serves to renormalize upward the longitudinal sound velocity. This latter effect is already accommodated by using the correct value of $v_l$ and thus no other effects of the scalar electron-phonon coupling, $S_\mathrm{(i)}$, remain. 

The only remaining term thus is the second -- quadrupole term in Eq.~(\ref{eq:Siii}), which coincides exactly with Eq.~(\ref{eq:S-u}). The latter was derived using phenomenological Schmid form, Eq.~(\ref{eq:electron-phonon}), of the electron-phonon coupling. The present first principles derivation provides thus an independent justification for the Schmid theory \cite{Schmid}.

\section{Ultrasonic attenuation}
\label{app:attenuation} 
\subsection{Normal metals}

For the sake of completeness we outline calculation of the ultrasonic attenuation. It is found by integrating out electronic degrees of freedom, $\hat Q$, and focusing on modification of the 
phonon propagator (\ref{eq:ph-propagator}) due to electron-phonon coupling. In the leading approximation it is given by the 
action (\ref{eq:S-u}), where one puts $\hat Q=\hat \Lambda$,  cf. Eq.~(\ref{eq:Lambda}), and integrates over the energy:
\begin{equation}
				\label{eq:S-Lambda-u}
iS_{\hat\Lambda,\mathbf{u}}\!=\! -\nu p_F^2\!\! \sum\limits_{\mathbf{q},\omega}
\bar{\mathbf{u}}^{\mu,\alpha}_{\mathbf{q},\omega}   \hat K^{\alpha\beta}(\omega) \frac{Dq^2}{d+2} 
\!\left [\delta_{\mu\nu}\! +\! \frac{d\!-\!2}{d} \frac{q_\mu q_\nu}{q^2} \!\right]\! \! \mathbf{u}^{\nu,\beta}_{\mathbf{q},\omega},  
\end{equation}
where the kernel is 
\begin{equation}
				\label{eq:K-kernel} 
\hat K^{\alpha\beta}(\omega)=\frac 1 4 \!\int\! d\e \, \mathrm{Tr}\left\{ \hat\gamma^{\alpha} \hat\gamma^{\beta}- \hat\Lambda_{\epsilon-\omega} \hat\gamma^{\alpha}  \hat\Lambda_{\epsilon}\hat\gamma^{\beta} \right\}.
\end{equation}
For normal metals one finds, cf. Eq.~(\ref{eq:Lambda}),  the dissipative Caldeira-Leggett kernel:
\begin{equation}
				\label{eq:Caldera-Leggett} 
\hat K(\omega)=\left(\begin{array}{cc} 0 & -\omega \\
\omega & 2\omega {\cal B}_\omega \end{array} \right) 
\end{equation}
and ${\cal B}_\omega=\int \! d\epsilon\, (1-F_{\e-\omega}F_\e)/(2\omega)$ is the bosonic distribution function. It provides damping to the phonon action, Eq.~(\ref{eq:phonon-action}): $(\omega\pm i0)^2 - \left(\omega_\mathbf{q}^{(j)}\right)^2\to 
\omega^2 \pm i\gamma_\mathbf{q}^{(j)} \omega - \left(\omega_\mathbf{q}^{(j)}\right)^2 $  (along with the fluctuation-dissipation related noise), where  the damping factors are \cite{Abrikosov,Shtyk,Shtyk-thesis}:
\begin{equation}
				\label{eq:damping}
\gamma_\mathbf{q}^{(j)} = c_j\, \frac{\nu p_F^2}{\rho_m}\, Dq^2, 
\end{equation}
where $c_t=2/(d+2)$ and $c_l=c_t[1+(d-2)/d]$. 

\subsection{Superconductors}

In the superconducting case the $\hat K(\omega)$ kernel, Eq.~(\ref{eq:K-kernel}), depends on the rotation angle in the Nambu space. For it's retarded, i.e. $\alpha=q$ and $\beta=cl$, component one finds:
\begin{eqnarray}
K^R(\omega)&=&-\frac 1 4 \int\!\! d\e\Big[  \cosh(\vartheta-\vartheta')F^L_{\epsilon'} - \cosh(\bar\vartheta-\bar\vartheta')F^L_{\epsilon} 
\nonumber \\ &+& \cosh(\vartheta-\bar\vartheta')(F^L_{\epsilon'}-F^L_\e)\Big], 
\end{eqnarray}
where $\e'=\e-\omega$. Since for $\e< \e_g$, $\mathrm{Im} \vartheta(\epsilon)=-\pi/2$ there is no contribution to the integral from the region $\e,\e'<\e_g$. The imaginary part of this expression, coming from the two energy intervals $\pm \e_g<\e<\pm \e_g+\omega$, is $\sim \omega^2$ at small $\omega$, which gives a small renormalization to the sound velocity. The real part, responsible for the attenuation, takes the following form:
\begin{equation}
						\label{eq:ReK}
\mathrm{Re} K^R(\omega) = \!\int\!\! d\e\,  \frac{\nu(\e)}{\nu} \frac{\nu(\e')}{\nu} \left[1 - 4u_\e v_\e u_{\e'} v_{\e'} \right]  (f_{\epsilon'}-f_\e),
\end{equation} 
where $2u_\e v_\e$ is given by Eq.~(\ref{eq:2uv}),   $f_\e=(1-F^L_\e)/2$ is the quasiparticle occupation number and the integral runs over the two intervals $\e<-\e_g$ and $\e> \e_g+\omega$, where both DOS are non-zero. 

For TRS superconductors the above expression takes the form:
\begin{equation}
				\label{eq:ReK-TRS}
\mathrm{Re} K^R(\omega)=  \int\!\! d\e\, \frac{\e\e'-\Delta^2}{\sqrt{\e^2-\Delta^2}\sqrt{\e'^2-\Delta^2}} \,(f_{\e'}- f_{\epsilon}).   
\end{equation}
In the case $\omega\ll \Delta,T$ one may expand over $\omega$ and use the fact that the fraction in Eq.~(\ref{eq:ReK-TRS}) tends to $1$ as $\omega\to 0$. As a result, $\mathrm{Re} K^R(\omega) = -2\omega \int_\Delta^\infty d\e \, (df_\e/d\e)=2\omega f_\Delta$.  This way the ultrasound attenuation coefficient in equilibrium disordered TRS superconductors is   found to be \cite{Tsuneto,Galperin,Abrikosov}:
\begin{equation}
				\label{eq:damping-super}
\gamma_\mathbf{q}^S = 2f\left(\frac{\Delta(T)}{T}\right) \, \gamma_\mathbf{q}^N,  
\end{equation}
where $f(\Delta/T)$ is the Fermi function.  
(The same relation holds in the clean case as well \cite{Gelikman}.)

It is worth noticing that the celebrated Eq.~(\ref{eq:damping-super}) does not hold in the TRS broken case. Indeed, in the limit $\omega\to 0$, the factor $\left(\frac{\nu(\e)}{\nu}\right)^2 [1 - (2u_\e v_\e)^2] =(\mathrm{Re} \cosh\vartheta)^2-(\mathrm{Re} \sinh\vartheta)^2\neq 1$ for complex $\vartheta(\e)$. In the small temperature case, $T< \gamma^{2/3}\Delta^{1/3}<\e_g$,   Eqs.~(\ref{eq:sqrtDOS}), (\ref{eq:2uv-1}) lead to: 
\begin{equation}
						\label{eq:ReK-1}
\mathrm{Re} K^R(\omega)\! =\! \frac{-4\omega}{3\gamma^{2/3}\Delta^{1/3}} \!\int\limits_{\e_g}^\infty \!\! d\e (\e-\e_g) \frac{df_\e}{d\e}= \frac{4\omega}{3\gamma^{2/3}\Delta^{1/3}} \!\int\limits_{\e_g}^\infty \!\! d\e f_\e,
\end{equation} 
In equilibrium this brings: 
 \begin{equation}
				\label{eq:damping-super-1}
\gamma_\mathbf{q}^S = \frac{4T}{3\gamma^{2/3}\Delta^{1/3}} \, e^{-\e_g/T} \, \gamma_\mathbf{q}^N,  
\end{equation}
i.e. there is an additional small factor $\sim T/(\gamma^{2/3}\Delta^{1/3})$ in comparison with TRS case (it may be 
overcompensated, though, by the fact that $\e_g<\Delta$).

\end{document}